%% file: arXiv_version.tex
\documentclass[12pt,
unsortedaddress,
 amsmath,amssymb,
 aps,
floatfix,
]{revtex4-2}
\usepackage[utf8]{inputenc}
\usepackage[caption=false]{subfig}
\usepackage[version=4]{mhchem}
\usepackage{siunitx}
\usepackage{xcolor}

\usepackage{listings}
\usepackage{epstopdf, epsfig}
\usepackage{graphicx}
\usepackage{dcolumn}
\usepackage{bm}

\newcommand{\AT}{{\alpha_{T/4}}}
\renewcommand{\textcolor}{}

\begin{document}

\preprint{APS/123-QED}

\title{Leading edge vortex formation and wake trajectory:\\ Synthesising measurements, analysis, and machine learning}

\author{Howon Lee}
\altaffiliation{Current Address: Dept of Aerospace Engineering, Georgia Tech, Atlanta GA}
\author{Nicholas Simone, Yunxing Su, Yuanhang Zhu}
\affiliation{Center for Fluid Mechanics, School of Engineering, Brown University, Providence RI}

\author{Bernardo Luiz R. Ribeiro, Jennifer A. Franck}
\affiliation{Department of Engineering Physics, University of Wisconsin-Madison, Madison, WI}%

\author{Kenneth Breuer}
\affiliation{Center for Fluid Mechanics, School of Engineering, Brown University, Providence RI}

\date{\today}

\begin{abstract}
The strength and trajectory of a leading edge vortex (LEV) formed by a pitching-heaving hydrofoil (chord $c$) is studied. The LEV is identified using the $Q$-criterion method, which is calculated from the 2D velocity field obtained from PIV measurements. The relative angle of attack at mid-stroke, $\AT$, proves to be an effective method of combining heave amplitude ($h_0/c$), pitch amplitude ($\theta_0$), and reduced frequency ($f^*$) into a single variable that predicts the maximum value of $Q$ over a wide range of operating conditions. Once the LEV separates from the foil, it travels downstream and rapidly weakens and diffuses. The downstream trajectory of the LEV has two characteristic shapes. At low values of $\AT$, it travels straight downstream after separating from the foil, while at higher values of $\AT$, an accompanying Trailing Edge Vortex (TEV) forms and the induced velocity generates a cross-stream component to the vortex trajectories. This behavior is accurately predicted using a potential flow model for the LEV and TEV. Supervised machine learning algorithms, namely Support Vector Regression and Gaussian Process Regression, are used to create regression models that predicts the vortex strength, shape and trajectory during growth and after separation. 
The regression model successfully captures the features of two vortex regimes observed at different values of $\AT$. However, the predicted LEV trajectories are somewhat smoother than observed in the experiments. The strengths of the vortex is often under-predicted. Both of these shortcomings may be attributed to the relatively small size of the training data set.
\end{abstract}

\maketitle

\section{Introduction}
The oscillating hydrofoil offers an appealing alternative to conventional rotary turbine hydrokinetic energy converters (HEC), with lower blade speeds, a low cut-in velocity, and a geometry favorable for shallow waters such as rivers and tidal estuaries \cite{Laws2016}. A typical configuration of the oscillating hydrofoil is shown in Fig.~\ref{fig:foilMotion}. The hydrofoil heaves upward with a high angle of attack inducing leading edge stall and the formation of a strong leading-edge vortex (LEV). The vortex core has an associated low pressure region acting on the upper surface of the foil which generates a large heaving force. Once the LEV separates from the leading edge and begins to convect downstream, the lift quickly deteriorates and the hydrofoil must rotate and heave in the other direction, repeating the energy harvesting cycle.

The first investigation of power extraction by a flapping foil was carried out by McKinney and DeLaurier \cite{McKinney1981}, demonstrating the potential to generate power from a steady flowing fluid. A numerical study by Kinsey and Dumas \cite{KinseyDumas2008} found that for a heaving amplitude of one chord length, efficiencies as high as 35\% could be obtained at a reduced frequency, $f^* = fc/U_\infty$, of 0.15 and maximum heaving amplitude of 0.75 chord length. Kim et al. \cite{Kim2017} divided the power generated by the foil into heave and pitch components which had different behaviors. The heave component of energy harvesting efficiency increased with reduced frequency. In contrast, the pitch component of efficiency decreased, with the efficiency even changing signs from positive to negative at times \cite{Kim2017}. Numerous field tests have also successfully demonstrated the energy harvesting capabilities in realistic environments \cite{Kinsey2011,Cardona2016, Mandre2015}.

Various studies have investigated methods to increase the energy harvesting efficiency of oscillating hydrofoils. Simpson et al. \cite{Simpson2008}, for example, showed experimentally that increasing the aspect ratio improves the efficiency of the foil. This agrees with the airfoil theory where the detrimental effect of the tip vortices on the lift generation capabilities of a foil weakens with greater aspect ratio. Kim et al. \cite{Kim2017} found that a similar relationship can be applied to the hydrofoil in heaving motion and observed that the change in efficiency is due to the power by heaving motion being significantly affected by the aspect ratio. It has been reported that efficiency increased when a trapezoidal pitching motion with a sinusoidal heaving motion was used rather than a sinusoidal motion for both heaving and pitching \cite{Ashraf2011,Young2014,Su2019_Resonant}. Kinsey and Dumas have also found that when end plates are mounted, the total efficiency is improved by reducing the 3D hydrodynamic losses \cite{Kinsey2012}. In a similar effect to that of increasing the aspect ratio, the benefits of end plates are usually thought to be the suppression of the tip vortex effects \cite{Kim2017}. However, with the increase in size of the end plate, the additional skin friction and flow structure interaction may work against this benefit, reducing the hydrodynamic force on the hydrofoil \cite{Kim2017}. Many studies have also explored another approach to improving energy harvesting efficiency by taking advantage of the wall confinement effect, operating the foil close to a wall inside a channel \cite{GarrettCummins2007,Su2019_confinement}. 

\textcolor{red}{Another practical means to improve the overall power production of an oscillating foil installation is to pack the foils into closely-spaced arrays. In the case of vertical-axis wind turbine arrays (VAWTs), arranging two counter-rotating VAWTs together can prove advantageous \cite{Dabiri2011} due to the ``constructive interference'' of the structured vortex wake shed by the leading turbine, which can improve the efficiency of the trailing turbine}. For oscillating hydrofoils, numerous experimental and computational studies have been conducted analyzing the performance of oscillating hydrofoils in tandem configuration. A 2D Unsteady Reynolds-Averaged Navier-Stokes (URANS) study by Kinsey and Dumas showed that having a counter-clockwise vortex above the trailing foil or a clockwise vortex slightly below the trailing foil has a favorable impact on power extraction \cite{Dumas2012}. Simeski and Franck \cite{Simeski2017} explored various combinations of vertical and horizontal separation between the foils and found that staggered configurations result in the efficiency of the trailing foil exceeding that of the leading hydrofoil. Ribeiro et al. \cite{Ribeiro2020} investigated the power extraction and LEV trajectory of an oscillating hydrofoil from large-eddy simulations (LES) and direct numerical simulations (DNS), where they identified two regimes of efficiency in terms of the relative angle of attack at mid-upstroke, or $\AT$. They found that for $\AT < 22^\circ$, the boundary layer remains attached to the foil with minimal separation and no distinct LEV while for $\AT > 22^\circ$, the LEV becomes more prominent and contributes to high efficiency \cite{Ribeiro2020}. 

\textcolor{red}{However, a taxonomy of the trajectory and strength of the LEV shed by the leading turbine in an array has not been completed over any range of operating parameters (frequency, pitch angle, and heave amplitude), and the lack of this knowledge, coupled with the difficulty in conducting experiments and computations of multi-foil configurations has inhibited progress in designing oscillating foil farms. We aim to fill this this gap in our knowledge in this work.}

This manuscript combines experimental measurements of oscillating foil turbine wakes with simple theory and machine learning tools to report on the strength, size, shape and trajectory of the LEV formed behind an oscillating hydrofoil. Particle Image Velocimetry (PIV) experiments are performed to obtain the LEV characteristics and position over a wide range of pitch and heave amplitudes and oscillating frequencies. Qualitative and quantitative analysis is used to explain the trends observed in the characteristics of the LEV and its wake trajectory after it is shed from the foil. 

\textcolor{red}{The PIV measurements generate vast quantities of data - unsteady velocity fields - and in order to take advantage of these fields in ways that we cannot intuitively guess, we also report on the use of machine learning tools to predict the strength, shape and trajectory of the LEV wakes. These tools, if accurate (and we will demonstrate that they are promising), provide a unique ability to predict LEV behavior for parameter combinations that we have not tested and still do not have complete modelling capability. The ML tools, once trained, will be cheap to use in guiding future design of oscillating flow turbine arrays.}

\textcolor{red}{The paper is organized as follows. In the following section we describe the techniques used in the measurements and review the methods used for vortex identification and machine learning (ML) training. In Section III, we discuss the experimental results, including the scaling of the vortex characteristics, as well as the success of a simple potential flow model used to predict the vortex trajectory after separation. The accuracy and limitations of the ML modelling is presented and discussed in Section II-C. Some concluding remarks complete the paper in Section IV.}

\section{Experimental Methods}
\subsection{Experimental Setup}
The experiments were conducted in the free-surface water flume at Brown University, with a 0.8 m wide and 0.6 m deep test section. The testing configuration is largely the same as that of Su et al. \cite{Su2019_thesis,ZhuSuBreuer2020} and Ribeiro et al. \cite{Ribeiro2021}, and consists of a single vertically-mounted hydrofoil with an elliptical cross section, with a chord length of 10 cm, and an aspect ratio of 3.5. End plates are mounted at each end of the foil in order to minimize effects from tip vortices. 

Particle Image Velocimetry (PIV) was performed to obtain the flow field around the hydrofoil and in the wake. The PIV setup is similar to that described by Su \cite{Su2019_thesis,ZhuSuBreuer2020}. The laser sheet was generated by a double-pulse laser (200 mJ Nd:YAG, EverGreen, Quantel USA, MT) with a wavelength ($\lambda$) of 532 nm. The flow was seeded with silver-coated hollow ceramic spheres (diameter: 100 $\mu m$, Potter Industries). Four Imager sCMOS cameras with 35 mm lenses were used to record the flow images at 15 Hz and the flow fields were calculated using Davis (v10, LaVision). A slightly modified experimental set up was used to obtain a second set of PIV data, used to verify the machine learning predictions. Fig. \ref{fig:setup} shows the camera configuration for these experiments, which used a single sCMOS camera (4 MegaPixels), capturing the flow field with the aid of a mirror positioned below the water flume, angled at {$45^\circ$}.

The hydrofoil can execute computer-controlled heaving and pitching motions. A linear servo motors (AeroTech) is used for the heaving motion and a stepper motor (Applied Motion Products) for the pitching motion. The pitch axis of the foil is located at the mid-chord. The pitching and heaving motions are described by:
\begin{equation}
\theta(t) = \theta_0 \sin (2\pi f t + \phi)
\label{eqn:pitching}
\end{equation}
\begin{equation}
h(t) = h_0 \sin (2\pi f t)
\label{eqn:heaving}
\end{equation}
where $\theta_0$ and $h_0$ are maximum pitching and heaving amplitudes respectively, $f$ is the oscillation frequency, and $\phi$ is the phase difference between the two motions. A phase difference of $\phi = \pi/2$ was used for all experiments, the value for the optimal energy harvesting performance \cite{McKinney1981}. The effective angle-of-attack of the foil at mid-stroke ($t/T = 0.25$), $\AT$ is defined as
\begin{equation}
\AT = \tan^{-1}(-2\pi~h_0/c~fc/U_\infty) + \theta_0\ ,
\label{eqn:alphaT4}
\end{equation}
and has been shown \cite{Kim2017,Ribeiro2021} to be a useful parameter to describe the overall energy-harvesting efficiency of the hydrofoil.

Table \ref{table:parameters} shows the range of foil kinematics measured with the PIV experiments. They were conducted at three different reduced frequencies, $f^* = fc/U_{\infty} = 0.08, 0.12$, and 0.15. Pitching and heaving amplitudes were varied when $f^*$ was fixed at 0.12. The pitching amplitude, $\theta_0$, was varied from $55^\circ$ to $85^\circ$ in increments of $10^\circ$, while the heaving amplitude, $h_0/c$, was varied from 0.5 to 1.25 in increments of 0.25. At other frequencies, the pitching and heaving amplitudes were fixed at $65^\circ$ and 1.0, respectively. These parameter combinations resulted in a variation of $\AT$ ranging from 0.31 to 0.84.

\subsection{Vortex Tracking}
The oscillating flow turbine is characterized by the formation, growth, separation and advection of a strong leading-edge vortex (LEV) \cite{KinseyDumas2008,Kim2017}. A typical velocity field, obtained from PIV measurements is shown in Fig.~\ref{fig:vorticityPlot}. The LEV was detected and tracked using the $Q$-criterion \cite{Hussain1995}. The gradient of the velocity field, $\nabla \bm{\mathrm{v}}(\bm{x},t)$, can be decomposed into the sum of the rate of vorticity tensor and the rate of strain tensor: $\nabla \bm{\mathrm{v}} = \bm{\Omega} + \bm{S}$, where 
\begin{equation}
  \bm{\Omega} = \frac{1}{2}[\nabla\bm{\mathrm{v}} + (\nabla\bm{\mathrm{v}})^T], 
\end{equation}
and 
\begin{equation}
  \bm{S} = \frac{1}{2}[\nabla\bm{\mathrm{v}} - (\nabla\bm{\mathrm{v}})^T].
\end{equation}
\noindent Here, $T$ denotes the transpose of a matrix.

The $Q$-value is defined as
\begin{equation} 
Q = \frac{1}{2}[|\bm{\Omega}|^2 - |\bm{S}|^2], 
\label{eqn:Qcriterion}
\end{equation}
and a ``vortex'' is identified as regions where the norm of the rate of vorticity tensor is greater than that of the rate of strain tensor, $Q > 0$. For this study, an interrogation window was drawn manually around the vortex observed from the PIV field data in order to fully capture the LEV in each frame. The size of the window is dependent on the kinematics of the foil which affects the size of the LEV. It, however, does not exceed beyond a chord length. This agrees with the findings of Rival et al., \cite{Rival2014}, where they discovered that the critical LEV diameter is one chord length. $Q$-values were then calculated from the velocity field (eq. \ref{eqn:Qcriterion}). The centroid of the maximum 300 $Q$-values was then used to define the position of the LEV core (Fig~\ref{fig:QPlot}). For each vortex, an ellipse with the same image moment (i.e. the equivalent moment ellipse) was fitted. The lengths of the two semi-axes of the ellipse was used to define the vortex size, shape, and orientation.

\subsection{Machine Learning}
\textcolor{red}{
The Machine Learning Toolbox (MATLAB, Mathworks, Natick MA) was used to train machine learning regression models that predict vortex characteristics and behavior. Several machine learning algorithms were evaluated before settling on Gaussian Process Regression (GPR) and Support Vector Machines (SVM) for use in this work. These were chosen for their robustness against outliers and overall flexibility \cite{Awad2015}. Gaussian process regression (GPR) models are nonparametric kernel-based probabilistic models \cite{MATLABmanual}. Gaussian processes (GPs) develop these kernels adaptively based on the available data, and provide probability distributions for the respective model parameters. GPs have been used extensively in time-dependent problems, which makes them appealing for our study of LEV behavior over time. Furthermore, with a proper choice of kernel function, the prediction capabilities of a GPR model can be enhanced. As we have observed exponential behavior in the physics of an LEV (such as vortex strength decay), the exponential kernel function was chosen. However, it should be noted that GPR can be computationally expensive which could be problematic with increasing data size \cite{Brunton2020}.}

\textcolor{red}{While the exponential GPR is proved to perform well for many of our parameters, it does not handle discontinuities very well \cite{Zhang2018}. This is problematic for the vortex trajectory predictions which exhibit rapid changes in direction when they separate from the hydrofoil (as discussed in later sections). For this reason, Support Vector Regression (SVR) was used to predict the $x$ and $y$ position of the LEV. SVR is an extension of Support Vector Machine (SVM) - a popular machine learning algorithm used for classification \cite{Vapnik}. In the field of fluid mechanics, SVR has been successfully applied to turbulence modeling and reduced-order modeling \cite{Fukami}. The implementation employed in the current work is the linear epsilon-insensitive SVM ($\epsilon$-SVM) regression. Here, the value of $\epsilon$ defines a margin of tolerance where no penalty is given to errors. In $\epsilon$-SVM regression, the set of training data includes the velocity fields as well as several predictor variables and the observed response values. The training goal is to find a function that deviates from the ground truth, by a value no greater than a specified tolerance $\epsilon$ for each training point $x$, and at the same time is as flat as possible and thus less sensitive to perturbations in the features \cite{MATLABmanual}. The user can tune $\epsilon$ against noise through a ``loss function'', which balances the various learning objectives (e.g., accuracy, simplicity, smoothness, etc.) \cite{Brunton2020}. SVM algorithms use a set kernel function, which in this case we choose a third-order polynomial (``cubic''). SVM has the advantage that it is capable of maintaining higher precision in the case of nonlinearity and small samples \cite{Liu2020}, thus amenable to the small data set of non-phase-averaged velocity fields. It is also computationally inexpensive in comparison to the GPR.}

\subsubsection{Training Process}
\textcolor{red}{A separate training was performed for each of five different response values: the $x$ and $y$ position of the vortex, its size (defined by the the lengths of the semi-axes of its equivalent moment ellispe), orientation, and $Q$-value. For each training performed, a total of 844 experimentally measured velocity fields, each with two components of velocity, $(u,v)$, were used as the basis of the training data, sampled from the range of parameters identified in Table \ref{table:parameters} and at different times, $t/T$, during the pitch-heave cycle. A mix of phase-averaged and instantaneous flow fields were used depending on data availability (see Table \ref{table:parameters}). Each velocity field in the training set also included the appropriate response value as well as several other predictor values: the relative angle of attack at mid-stroke, $\AT$, the nondimensional frequency, $f^*$, the heaving and pitching positions, $h(t), \theta(t)$, and the nondimensional cycle time, $t/T$. All values were appropriately normalized by freestream velocity, chord, frequency. The $Q$-value was normalized by its maximum value for that specific set of kinematics.}  

\textcolor{red}{A good supervised machine learning model should be generalizable, providing good predictions from previously unseen data. In this respect, cross-validating the model prevents overfitting and prevents the model to fit the training data perfectly at the cost of its generalizability and real life applications \cite{Brunton2020}. In this study, the selected scheme of cross-validation was ten-fold cross-validation. The training data was partitioned so that 30 $\%$ of the data was also used for validation.}

The prime metric for assessing the performance of the trained model in predicting a generic quantity, $a$, is the root-mean-square error:
\begin{equation}
RMSE_a = \sqrt{\sum_{i=1}^{n} \frac{(\hat{a_i}-a_i)^2}{n}}
\label{eqn:RMSE}
\end{equation}
where $n$ is the size of the training data set (in this study, 844 fields). $\hat{a_i}$ are the predicted values of the variable $a$ generated by the machine learning process - for example, the $x$-location of the LEV, its normalized $Q$-strength, etc. In contrast, $a_i$ are the ground truth data.

\section{Results and Discussion}
\subsection{LEV Behavior}

A qualitative view of the vortex core, represented by the $Q$-value ``cloud'' (Fig~\ref{fig:LEV_vs_Time}), clearly shows the LEV growing in strength, as the increase in the red area of the LEV from $t/T = 0.26$ to $t/T = 0.44$ indicates. The LEV then separates from the foil and dissipates in the wake as it travels downstream. The evolution of the vortex strength was quantified by taking the average of the highest 50 $Q$-values from the 300 points within the cloud for each frame in time (Fig~\ref{fig:maxQvsTime}). The top 50 $Q$-values rather than just one maximum point were chosen because it reduced some of the frame-to-frame fluctuations among the maximum average $Q$-values; increasing the quantity beyond 50 did not yield any noticeable improvement. The LEV is formed early in the cycle, and retains its high strength, as vorticity is continuously fed into the vortex from the feeding shear layer that connects the LEV to the leading edge of the hydrofoil. 

At about $t/T= 0.44$, as the foil nears the pitch reversal point ($t/T = 0.5$), the LEV separates (indicated by the red marker in Fig~\ref{fig:maxQvsTime}), and advects downstream. After separation, the vortex begins to decay exponentially in strength (Fig~\ref{fig:maxQvsTime}). For the example shown, $\bar{Q}_{max} \sim \exp(-18.8 t/T)$.

The evolution of the size of the LEV also follows a characteristic pattern. From the cloud of $Q$ points, ellipses that have the same position and image moments \cite{Rocha2002,Candelier2016} are fitted. A typical example of the evolution of the LEV size is shown in Fig. \ref{fig:maxQvsTime}. Confirming our qualitative assessment (Fig~\ref{fig:LEV_vs_Time}) we see that the vortex remains as a stronger and compact structure until the point of separation (marked in red). After LEV separates from the foil, new vorticity is no longer being fed into the vortex and the vortex strength starts to decay. At the same time, we see a rapid increase in the area of the equivalent moment ellipse. 
Both the amplitude decay and area increase are due to a combination of turbulent dissipation and 3D mixing by the tip vortex.

Although the kinematics are defined by three parameters: the non-dimensional frequency of oscillation, $f^*$, heaving amplitude, $h_0/c$, and pitching amplitude, $\theta_0$, the maximum strength of the LEV, as measured by the highest $\bar{Q}_{max}$ throughout the cycle (i.e. $\max(\bar{Q}_{max})$), is well predicted by the relative angle of attack at mid-upstroke, $\AT$ (eq.~\ref{eqn:alphaT4}), and shows a monotonic rise over the range of $\AT$ considered (Fig~\ref{fig:maxQvsAlpha}). A purely empirical fit of this behavior is given by: $\max(\bar{Q}_{max}) = 377.51 \log(\AT) + 584.29$. Similar results were found by Ribeiro et al. \cite{Ribeiro2021}, who found from both PIV measurements and DNS simulations that, with few exceptions, the strength of the primary leading-edge vortex increases with increasing relative angle of attack.

\subsection{Behavior of the LEV trajectory in the wake}
\subsubsection{Experimental observations}
Using the $Q$-criterion method of vortex detection, we identify the trajectory that the LEV follows in the PIV field of view - approximately 2 to 2.5 chord lengths downstream from the foil. Two regimes of trajectories were identified. The LEV trajectories for values of $\AT$ below $\sim 0.5$ can be seen in Fig.~\ref{fig:regime1Traj} and will be denoted as the ``LEV regime" following the nomenclature by Ribeiro et al. \cite{Ribeiro2021}. In this regime, the LEV initially remains attached to leading edge as the foil heaves upwards and follows its motion. It detaches from the foil soon after the foil reaches its maximum heave amplitude, subsequently advecting downstream with minimal $y$-displacement, which results in this regime's characteristic ``hockey stick" trajectory. 

At higher values of $\AT$ greater than 0.49, the trend in trajectory changes (Fig \ref{fig:regime2Traj}). This regime is denoted as ``LEV+TEV regime" \cite{Ribeiro2021}, and is characterized by the presence of an additional vortex, of opposite sign, in the flow field. Depending on the foil kinematics, this new vortex is formed either due to separation at the trailing edge which creates a trailing-edge vortex (TEV), or by the vorticity sheet that forms below the LEV on the upper surface of the foil consolidating into a vortex at the trailing edge and shed into the wake. In the LEV+TEV regime, the LEVs share a similar initial trajectory when the foil remains attached to the leading edge. After separation, however, the vortex exhibits a positive $y$-velocity and forms a curved trajectory as it travels downstream.

The straight downstream trajectory of the separated LEV for low values of $\AT$ are simply due to the fact that the vortices travel with the local flow, and the local flow is predominantly in the $x$-direction. In contrast, the cause of the positive $y$-velocity in the LEV+TEV regime is likely a consequence of the presence of the additional vortex which interacts with the LEV. The counter-clockwise vortex's counter-clockwise rotation induces the upwards motion of the LEV, while the clockwise rotation of the clockwise vortex induces the same effect on the TEV. The LEV trajectory resultant from this interaction is illustrated schematically in Fig.~\ref{fig:summaryLEVTEVRegime}.

\subsubsection{Potential Flow Model}
In order to verify this explanation of the vortex trajectories observed in the LEV+TEV regime, a simple model based on potential flow theory was tested. The LEV and TEV each induce a velocity on the other: $u_{i} = \frac{\Gamma_j}{4\pi d}$ where $\Gamma_j$ is the circulation and $2d$ is the distance between the vortex cores. The circulation is obtained by taking a contour integral of the velocity field from the PIV measurements.

The circulation of the LEV and TEV were assumed to be constant, established at the time the vortices separated from the hydrofoil. The position of each vortex, the vortex separation and the respective induced velocity vectors were updated in time using a MATLAB script to generate a predicted trajectory for the vortex pair. In order to incorporate the sensitivity of the trajectory predictions to uncertainties in the initial vortex positions and strengths, a Monte Carlo method was used, where the position of the contour integral and the vortex location was varied by up to 0.1c, and a total of 100 trajectories were generated and averaged to obtain a mean trajectory and uncertainty limit. Viscous decay was ignored in this model.

An example of the result from this simulation is shown in Fig.~\ref{fig:experimentalPotential}. The predictions based on potential flow theory agree well with the experimental data, capturing the upward motion of the LEV that is characteristic of LEV+TEV regime. It should be noted, however, that the three dimensional effects from tip vortex contribute to the experimental vortex trajectory which results a decrease in accuracy of the two-vortex potential flow model. However, the potential flow model was also applied to CFD data of oscillating foils' vortex wake \cite{Ribeiro2020} for which longer downstream evolutions were available. The model also demonstrated excellent agreement over a range of operating conditions (Fig \ref{fig:PotentialFlowModel}).

In cases where the model prediction was less faithful to the observations, largely two types of deviations can be observed. The first is when the predicted initial slope of the upwards motion was not as steep as the CFD data indicated, (e.g. Fig.~\ref{fig:PotentialD}). This error can be attributed to the sensitivity of the model to the vortex circulation. The second deviation between the model and the observations were increasing errors as the LEV traveled downstream (e.g. Fig.~ \ref{fig:PotentialC}) shows that the predicted trajectory begins to flatten at $x/c = 3$ while the vortex predicted by the CFD data travels further down in y-direction. The CFD data shows additional vortex formations that occur which the potential flow model does not take into account. The resultant interactions between numerous vortices likely causes these differences. 

In summary, the LEV trajectory characteristics of the LEV+TEV regime is due to the introduction of a vortex of opposite sign. The direction of the velocity induced by the new vortex depends on the position of the TEV with respect to the LEV. When the vortex pair is initially shed from the foil, there is a net positive $y$-velocity. Because the circulation of the LEV is greater than that of the TEV, the TEV moves faster and orbits about the LEV. When the TEV has rotated more than 90 degrees, a net negative velocity is induced. Therefore, a downward concave curve is observed in the experiments. After the TEV rotates 270 degrees about the LEV, the induced velocity is in the positive $y$-direction once again which flattens out the trajectory. This occurs at approximately $x/c \sim 3 - 4$. Beyond that point, the same interaction will eventually cause a change in direction within the trajectory.

\subsection{Machine Learning}

After the training process outlined in Section 2, the models that provided the optimal results are outlined in Table \ref{table:models} which includes the quantitative errors. The training times were attained using Intel Core i7-6700HQ CPU and GeForce GTX 960M GPU.

A qualitative assessment of the success of the machine learning process is the ``Predicted vs. Actual'' plot \cite{MATLABmanual,Brink2017}, the results of which are shown in Fig.~ \ref{fig:PredictedvsActual} for each of the quantities tested. A high performance model should have points clustered along the 1:1 diagonal, with small deviations. If any clear patterns, different from the perfect prediction diagonal, are observed in the plot, it is likely that the model can be improved and different types of models can be explored to ensure the most optimal results. 

Overall, it can be seen that the regression model predictions of the LEV positions (Figs. \ref{fig:xPrediction}, \ref{fig:yPrediction}) agree very well with training data, as demonstrated by the points being clustered around the 1:1 line. Note that the horizontal bands in each of the Predicted-vs-Actual plot reflect the fact that at each value of $t/T$, the model predicts a single value for the $(x,y)$ position of the LEV, but uses multiple individual realizations as part of the training data.  

The shape of the best-fit ellipse (Figs.~\ref{fig:axisAPrediction}, \ref{fig:axisBPrediction}), is also well predicted by the regression, although not as successfully as the position prediction. The ``rounder'' appearance of the point cluster indicates that the predictions are less accurate when the LEV is mid-sized, and more accurate at the early and late stages of its evolution.

\textcolor{red}{
The prediction of the $Q$-value (Fig.~\ref{fig:qPrediction}) exhibits the worst performance, particularly for high values of Q where the ML consistently underpredicts $Q$, particularly for the mid-range values of $\AT$. This weakness corresponds to early in the vortex shedding cycle where the Q-value is high, but interestingly, the prediction is much better as the vortex weakens, corresponding to the later times in the cycle. From the perspective of developing a tool to predict the location and strength of shed vortices for tandem foil vortex interactions, this is encouraging performance. Although the maximum Q value of the vortex is useful, it is more important to accurately estimate the position and strength of the shed vortex further downstream at the point where it will interact with a second foil, or downstream object.
}

As long as the parameters remain similar to the range of the parameters used in the training data (Table \ref{table:parameters}), we should be able to use the machine learning results to predict the trajectory of an LEV formed from a oscillating hydrofoil. To reassure model's performance, a test parameter combination within the training data is used and a similar vortex tracking is observed between measured and predicted (Fig \ref{fig:trainingExample}).

However, the power of any machine learning utility is to compare predicted and observed LEV trajectories for parameter combinations that are not part of the training data. Qualitatively, a high performance model should accurately capture the (i) the sharp increase in the $y$-position early in the cycle, reflective of the movement of the foil, (ii) the development of a high $Q$-value during this early stage where the LEV is still attached to the foil, (iii) a change in behavior after separation, determined by the trajectory regime which is dependent on $\AT$, (iv) an increase in vortex size during this stage, reflective of the diffusion of the vortex, and lastly, (v) a decrease in the $Q$-value after separation from the foil, reflecting the vortex decay. Four parameter combinations, detailed in Table \ref{table:verificationParameters}, were used to test the accuracy of the ML regression. Two cases are at the nondimensional frequency $f^* = 0.12$, which was a common frequency in the training data, while the other two cases lie well outside the training range.

Fig. \ref{fig:MLshowcase} shows a side-by-side comparison between the measured trajectory obtained from PIV data analysis (left) and the corresponding machine learning prediction (right). At first glance, it can be seen that the predicted and true vortex characteristics agree with each other. The time-dependent traits, such as a decrease in vortex strength, represented by the colors of the vortices, and the increase in the size of the vortices, represented by the size of the equivalent ellipses, are accurately captured. The change in the $y$-velocity of the vortex between the LEV and the LEV+TEV regime is also reasonably well captured.

The differences between the observed and predicted behavior is instructive. The regression machine learning models tend to predict trajectories that are smoother, and the abrupt changes in direction - the ``hockey-stick'' trajectories observed at low $\AT$ are less pronounced in the predictions, replaced by smoother paths. An example of this is the trajectory of the LEVs shown in Figs~\ref{fig:e} and \ref{fig:f}. In the PIV data, at approximately $x/c = 0.3$, the LEV abruptly changes course once separation from the foil occurs. The machine learning model's interpretation of this behavior qualitatively agrees, but is much more gradual, resulting in a smoother trajectory.

\textcolor{red}{
A point to concern when selecting training parameters is the possible coupling between different input variables which, if present, may reduce the accuracy of the predictions if not taken into account. For example, in a limited study, a single model was trained to simultaneously predict both the $x-$ and $y-$position of the LEV. This model yielded similar, but at times, poorer, predictions compared to those generated by the independent models. The lack of improvement may come from the fact the change in the $x$-position is dominated by the freestream velocity, which is significantly greater than the induced velocity in the $x$-direction from the vortex pair interaction. In contrast, the $y$-trajectory is strongly affected by the induced velocity of the second vortex.
}

Another distinct shortcoming in the machine learning performance is in the prediction of the LEV amplitude. This is already reflected in the predicted-actual data (Fig.~\ref{fig:qPrediction}), and is confirmed here. Observing the development of the maximum normalized $Q$ value through time in the experimental data in Fig. \ref{fig:MLshowcase}, we see that the LEV strength decays relatively slowly, retaining a high $Q$-value greater than 0.5 early in the cycle, until the vortex convects to approximately $x/c = 1$. The higher normalized $Q$-values are indicative of the LEV remaining attached to the leading edge of the foil. It is only after separation that the LEV begins to decay at significantly higher rate (e.g. Fig \ref{fig:e}). This behavior is not captured with the same accuracy by the machine learning predictions, which observe a more gradual decrease in the $Q$-value (Fig \ref{fig:f}) after separation. The consequence of this is that more frames are required to achieve the same change in the rapid change in $Q$. In some cases, the overall decay of the vortex seems delayed, retaining $Q$-values near 1 until the point of separation. This issue can likely be resolved with a greater number of training data. 

Despite these differences, the machine learning predictions for the LEV trajectories are remarkably accurate, and a quantitative analysis of the errors is shown in Fig. \ref{fig:errors}. Most of the time, the absolute position error (the distance between the actual and predicted LEV centroid positions) remains below 0.15c. As the trajectory evolves, the error accumulates, resulting in the highest overall errors at the larger times. The smoothing of the trajectory, mentioned above, also contributes to the error. In particular, between $t/T \sim 0.3 - 0.4$, and the LEV separates from the foil, the experimental trajectory experiences a sharp change in direction. As the machine learning equivalent smooths out the abrupt shift, the errors during this time period are larger, confirmed by the bumps observed for all four validation cases in Fig. \ref{fig:errors}. 

\textcolor{red}{
Despite promising performance of the machine learning predictions, it can be seen in Fig. \ref{fig:errors} that the ML prediction errors are higher than those of the potential flow model predictions, which are based on a physical analysis of the experimental data. However, it should be noted that the potential flow model is initiated at the point of separation with the inputs of the LEV-TEV vortex pair's respective positions and circulations. As such, it starts with the positional error of $0$ at a much later point in the cycle and an accurate measure of the key vortex circulation. Yet, we see that the rate of increase in the error is roughly similar to that of the machine learning model, indicating that the ML model performs as well as the potential flow model with regards to the propagation of the error. This, in addition to the fact that machine learning model is capable of generating a prediction from just the foil kinematic, means that the machine learning model could be an attractive alternative. Furthermore, once it is trained, the machine learning model produces its predictions in a much shorter time than the potential flow tool for forecasting the wake vortex topology without extensive measurements or further calculations.
}

\section{Concluding remarks}
The ability to accurately predict the formation, separation and downstream trajectory of LEVs over a range of operating parameters is of fundamental interest, as well as of practical utility. Several physical systems rely on an accurate prediction of the strength and location of LEVs, including understanding the dynamics of fish schooling \cite{Whittlesey2010}, as well as optimal placement within arrays of vertical axes wind turbines \cite{Dabiri2011} or oscillating hydrofoils \cite{Ribeiro2021}.

Here, a heaving-pitching hydrofoil is used to generate LEVs of varying strength, and in analysing PIV measurements over a wide range of parameters, we have found that, although there are three parameters needed to fully describe its motion - pitching, heaving amplitude and frequency - the effective angle of attack at midstroke, $\AT$, is a convenient quantity that collapses all three parameters and serves as an accurate predictor of subsequent LEV characteristics. 

The strength of the LEV, measured by the maximum $Q$-value, increases with $\AT$. In agreement with other results \cite{Ribeiro2020}, the trajectory that the LEV follows after separation can be loosely divided into two regimes: LEV and LEV+TEV regimes, where the transition occurs at approximately $\AT = 0.49$. In both cases, the LEV follows the motion of the hydrofoil until separation. For the LEV regime, the LEV simply convects downstream with minimal $y$-displacement. In contrast, in the LEV+TEV regime, an additional vortex with an opposite-signed strength is formed near the trailing edge. The presence of a vortex-pair results in a self-induced motion in the positive $y$-direction - a motion very well described by a simple potential flow model that relies only on the knowledge of the vortex strengths and positions as they separate from the hydrofoil.

\textcolor{red}{After gaining an understanding of the physics behind LEV trajectory, a machine learning approach was used to create a reduced-order modelling tool to capture trends that were found.} Supervised regression machine learning was found to accurately predict numerous LEV characteristics over a wide range of foil kinematic parameters. The trained model was successful in capturing the two vortex trajectory regimes, as well as the progression of the vortex size and strength, \textcolor{red}{particularly after the initial phase of vortex separation}. Verification of the regression model, using additional PIV data, indicates that the error in the predicted trajectory is small, usually limited to 0.15 chord lengths.

\textcolor{red}{Although there is no substitute for a detailed understanding the physics of LEV formation, separation, and advection, the machine learning (ML) tools provide the capability to generate useful predictions of the wake structure from just the foil kinematics without the need for time-consuming experiments or high-fidelity numerical simulations. The efficiency of the ML prediction is ideal for exploring a large parameter space required in optimization problems, and has the potential to be an attractive tool to rapidly reduce the size of the search space required to design arrays of oscillating hydrofoils that can harvest the energy from the LEV shed from upstream devices \cite{Dabiri2011,Ribeiro2020}. Of course the ML results have clear limitations. In the present study we have trained the predictions at a single Reynolds number, and with a limited set of kinematics that only considers sinusoidal motion. Reynolds number effects on single foil performance have been shown to be very modest \cite{Ribeiro2020}, although there is no data on Re-effects on the wake behavior. In addition, extrapolation of the model for parameters that stray far from the training data would be of questionable reliability. As such, more complex kinematics may well prove attractive \cite{Young2013, Fenercioglu2015} but at this stage, it is not yet clear whether $\AT$ will remain a good predictor of the leading foil performance \cite{Ribeiro2021} and how the ML predictions will fare as the kinematic space grows. These are clearly subjects for future study. Lastly, although these results show excellent promise, other deep learning techniques, such as convolutional neural networks \cite{Raissi2019} may prove to be an attractive alternative that might demonstrate better performance than is observed in this initial study.}

\section{Acknowledgements}
This research was funded by the US National Science Foundation (CBET award 1921594: JF,BR; CBET award 1921359: KB) and the US Air Force Office of Scientific Research (Grant FA9550-18-1-0322: KB, NS, YS, YZ). HL was also supported by the Karen T. Romer Undergraduate Teaching and Research Award from Brown University.

\clearpage
%
\input{arXiv_version.bbl}

\clearpage
\input{tables}

\clearpage
\input{figures}

\end{document}

%% file: arXiv_version.bbl
%

%% file: tables.tex
\begin{table}
\centering
\caption{The kinematic parameters of an oscillating hydrofoil explored in this study, the corresponding number of velocity field frames used for machine learning training matrix, and whether they were phase-averaged data. The kinematics are arranged in ascending order of $\AT$, the relative angle of attack at mid-stroke.}
\vspace{2mm}
\begin{tabular}{||c c c c c c||} 
 \hline
 $f^*$ & $\theta_0 (^{\circ})$ & $h_0/c$ & $\AT$ & $\#$ Frames & Phase Avg.\\ [0.5ex] 
 \hline\hline
 0.12 & 55 & 1 & 0.31 & 187 & O\\ 
 0.12 & 65 & 1.25 & 0.38 & 20 & X\\
 0.15 & 65 & 1 & 0.38 & 20 & X\\
 0.12 & 65 & 1 & 0.49 & 187 & O\\
 0.12 & 65 & 0.75 & 0.62 & 22 & X\\
 0.12 & 75 & 1 & 0.66 & 185 & O\\
 0.08 & 65 & 1 & 0.67 & 13 & X\\
 0.12 & 65 & 0.5 & 0.77 & 186 & O\\
 0.12 & 85 & 1 & 0.84 & 24 & X\\
 \hline
\end{tabular}
\label{table:parameters}
\end{table}

\begin{table}
\centering
\caption{Summary of models used for each feature predicting algorithm with their training times and root-mean-square errors (eq. \ref{eqn:RMSE}).}
\vspace{2mm}
\begin{tabular}{||c c c c||} 
 \hline
 Predicted feature & Model & Time (s) & RMSE \\ [0.5ex] 
 \hline\hline
 X position & Cubic SVM & 12.50 & 0.11 \\ 
 Y position & Cubic SVM & 17.49 & 0.072 \\
 $\bar{Q}_{max}/\max(\bar{Q}_{max})$ & Exponential GPR & 29.19 & 0.16 \\
 Ellipse major axis & Exponential GPR & 23.52 & 0.058 \\
 Ellipse minor axis & Exponential GPR & 46.14 & 0.038 \\
 \hline
\end{tabular}
\label{table:models}
\end{table}

\begin{table}
\centering
\caption{The kinematic parameters of an oscillating hydrofoil used for machine learning verification.}
\vspace{2mm}
\begin{tabular}{||c c c c c||} 
 \hline
 $\text{Case}$ & $f^*$ & $\theta_0 (^{\circ})$ & $h_0/c$ & $\AT$ \\ [0.5ex] 
 \hline\hline
 i & 0.12 & 80 & 1 & 0.75 \\ 
 ii & 0.10 & 75 & 1 & 0.75 \\
 iii & 0.12 & 70 & 1 & 0.58 \\
 iv & 0.10 & 55 & 1 & 0.40 \\

 \hline
\end{tabular}

\label{table:verificationParameters}
\end{table}

%% file: figures.tex
\clearpage \begin{figure}
\centering
\includegraphics[width=0.4\textwidth]{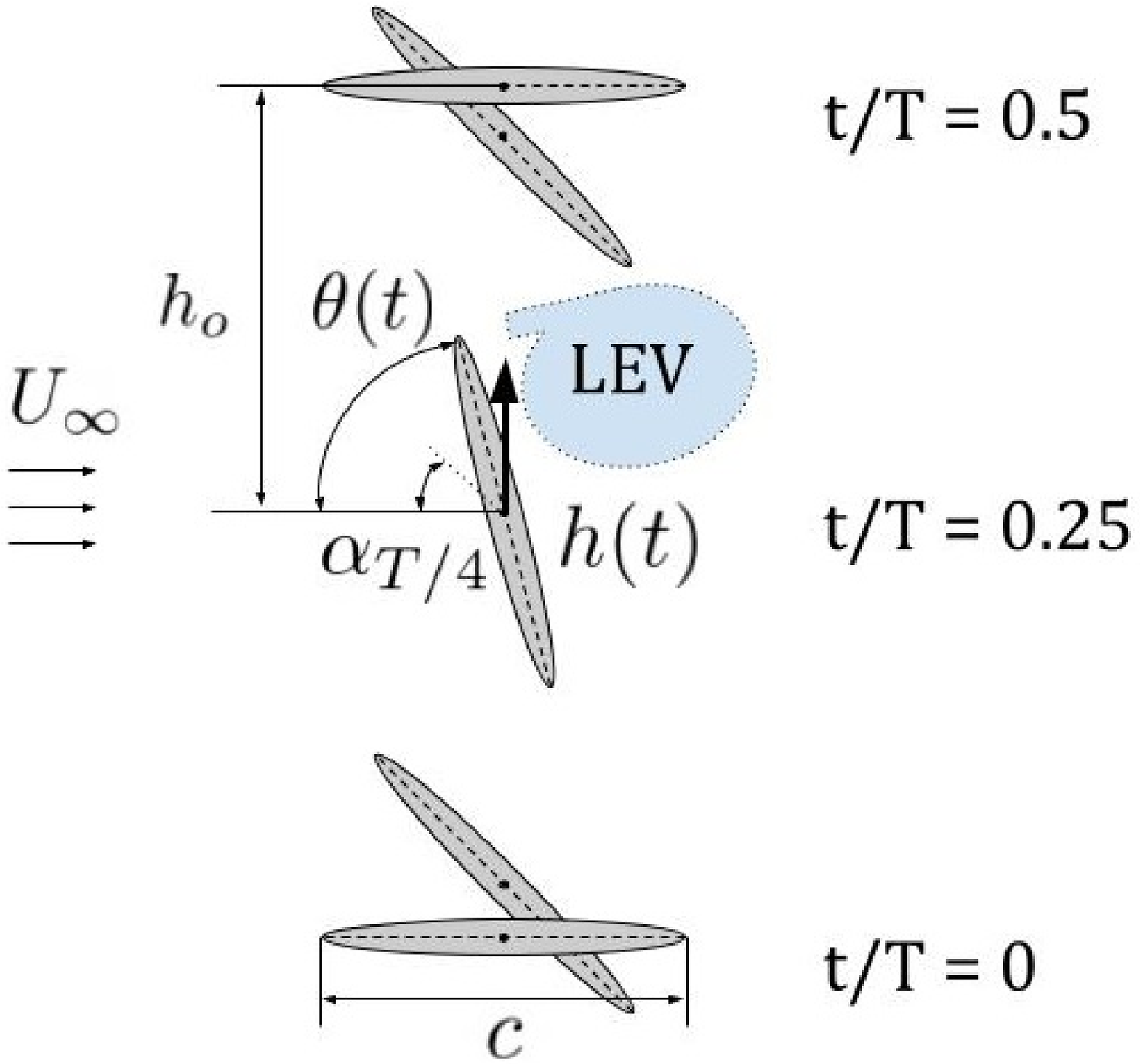}
\includegraphics[width=0.4\textwidth]{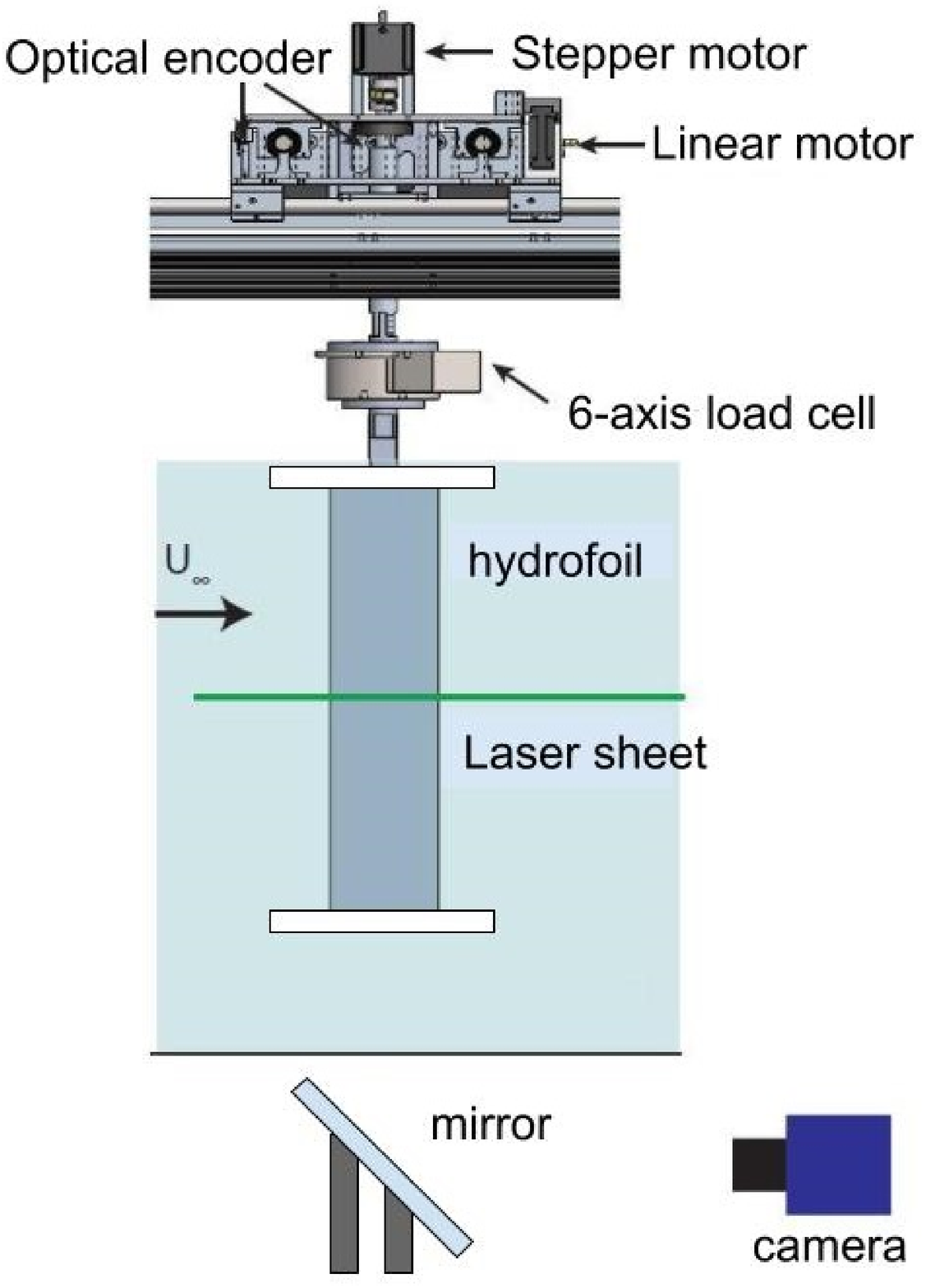}

\caption{Left: The motion of a single oscillating hydrofoil for a half-cycle from a top-down view. At the start of the cycle, $t/T = 0$, the foil is positioned at the negative peak amplitude and is angled at $0^\circ$. The foil follows a sinusoidal heaving and pitching motion with a phase difference of $\pi/2$. Right: The PIV experimental set up in the flume used for verification PIV data. 
}
\label{fig:foilMotion}
\label{fig:setup}
\end{figure}

\clearpage \begin{figure}
 \centering
 \subfloat[]{\includegraphics[width=0.4\textwidth]{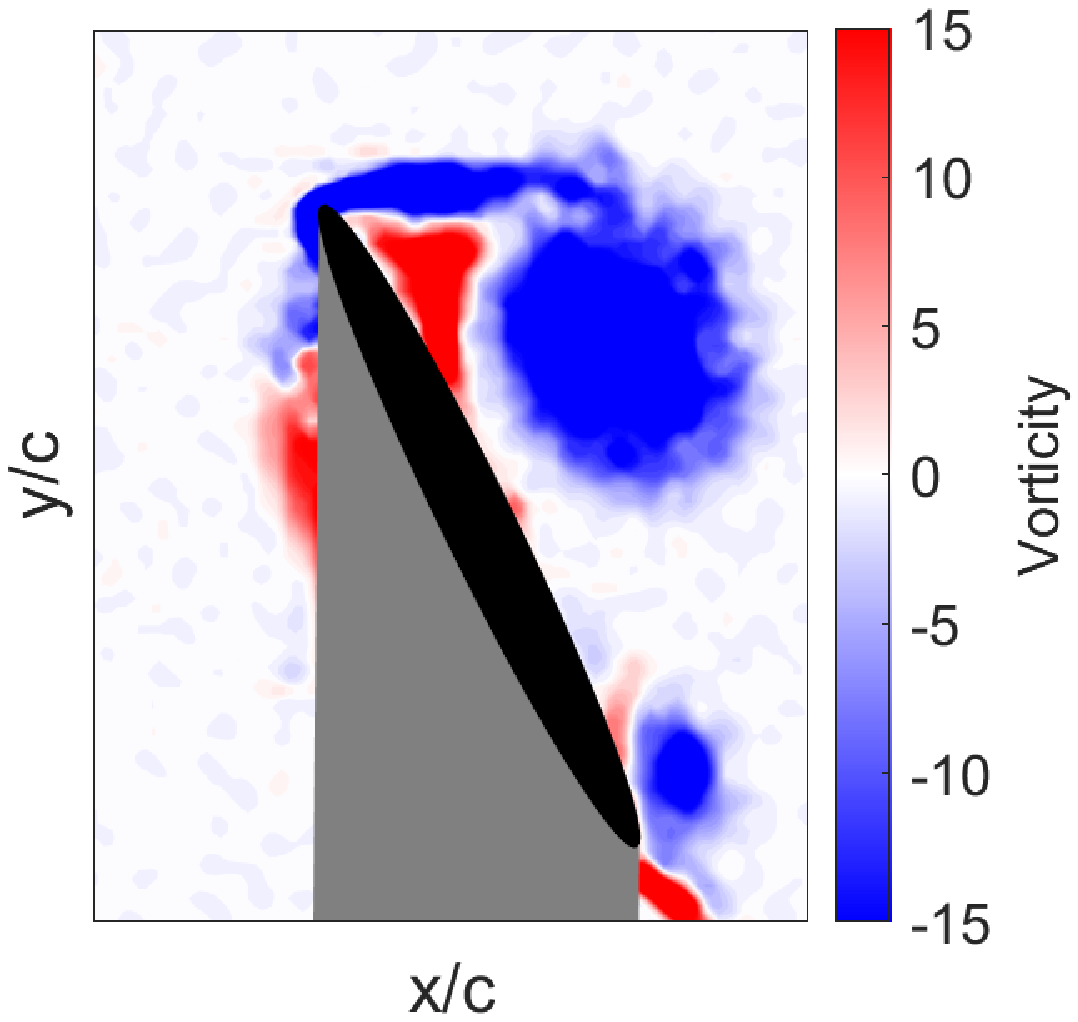}
 \label{fig:vorticityPlot}}
 \subfloat[]{\includegraphics[width=0.4\textwidth]{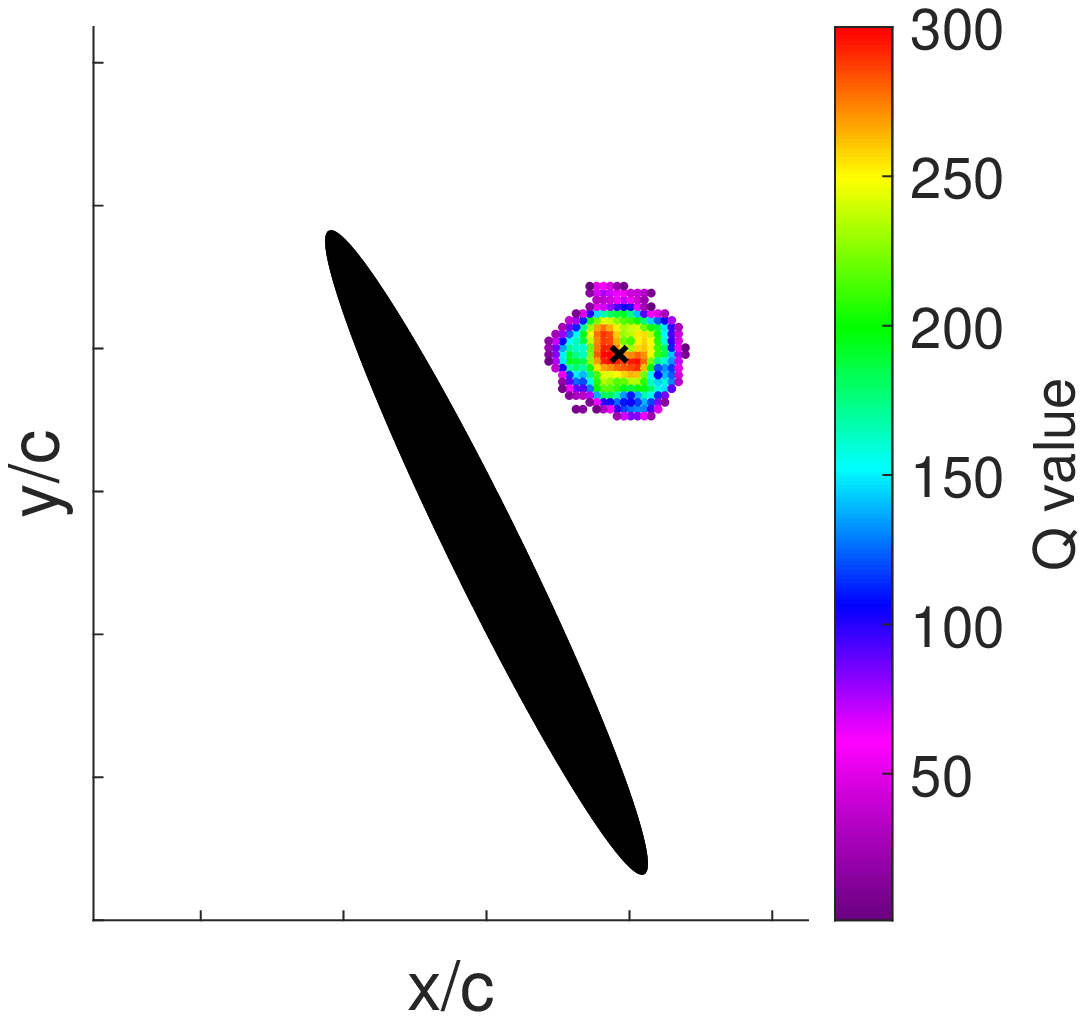}
 \label{fig:QPlot}}
 \caption{An example of $Q$-criterion used to detect and track LEV from the velocity field and its corresponding vorticity field. The foil's operating kinematics are: $f^* = 0.12, \theta_0 = 65^\circ, h_0/c = 0.5$ at $t/T = 0.30$. a) Nondimensionalized vorticity field for a typical case during mid-stroke from the PIV measurements. b) The corresponding ``cloud" of $Q$-values of the flow field. The black marker indicates the location of the LEV centroid. }
 \label{fig:QcriterionPlot}
 \end{figure}
 
\clearpage \begin{figure}
\centering
\includegraphics[width=0.6\textwidth]{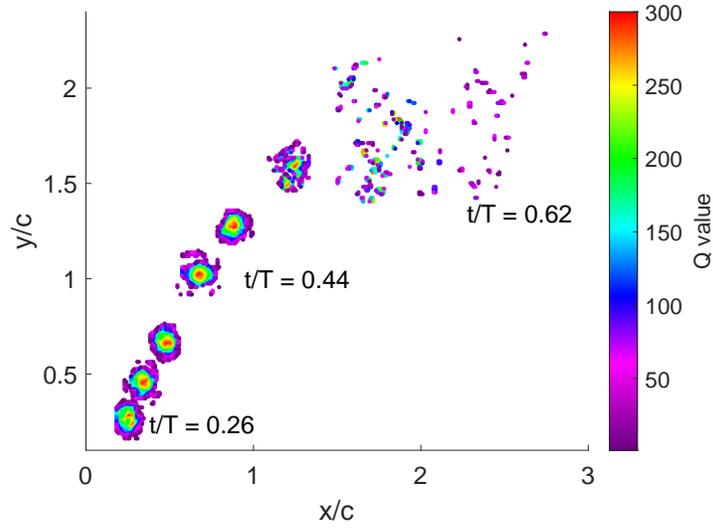}
\caption{An example of the evolution of the leading-edge vortex (LEV), as a function of time as indicated by the movement of clouds of $Q$-values. The largest 300 $Q$-values for each time are plotted. The tracking begins at $t/T = 0.26$ and ends at $t/T = 0.64$, while the clouds of $Q$-values are captured every 0.06 $t/T$. The growth, separation and advection of the LEV is typical over almost all parameter combinations. In this case, the kinematics are: $f^* = 0.12, \theta_0 = 85^\circ, h_0/c = 1$, $\AT = 0.84$.}
\label{fig:LEV_vs_Time}
\end{figure}

\clearpage \begin{figure}
\centering
\includegraphics[width=0.45\textwidth]{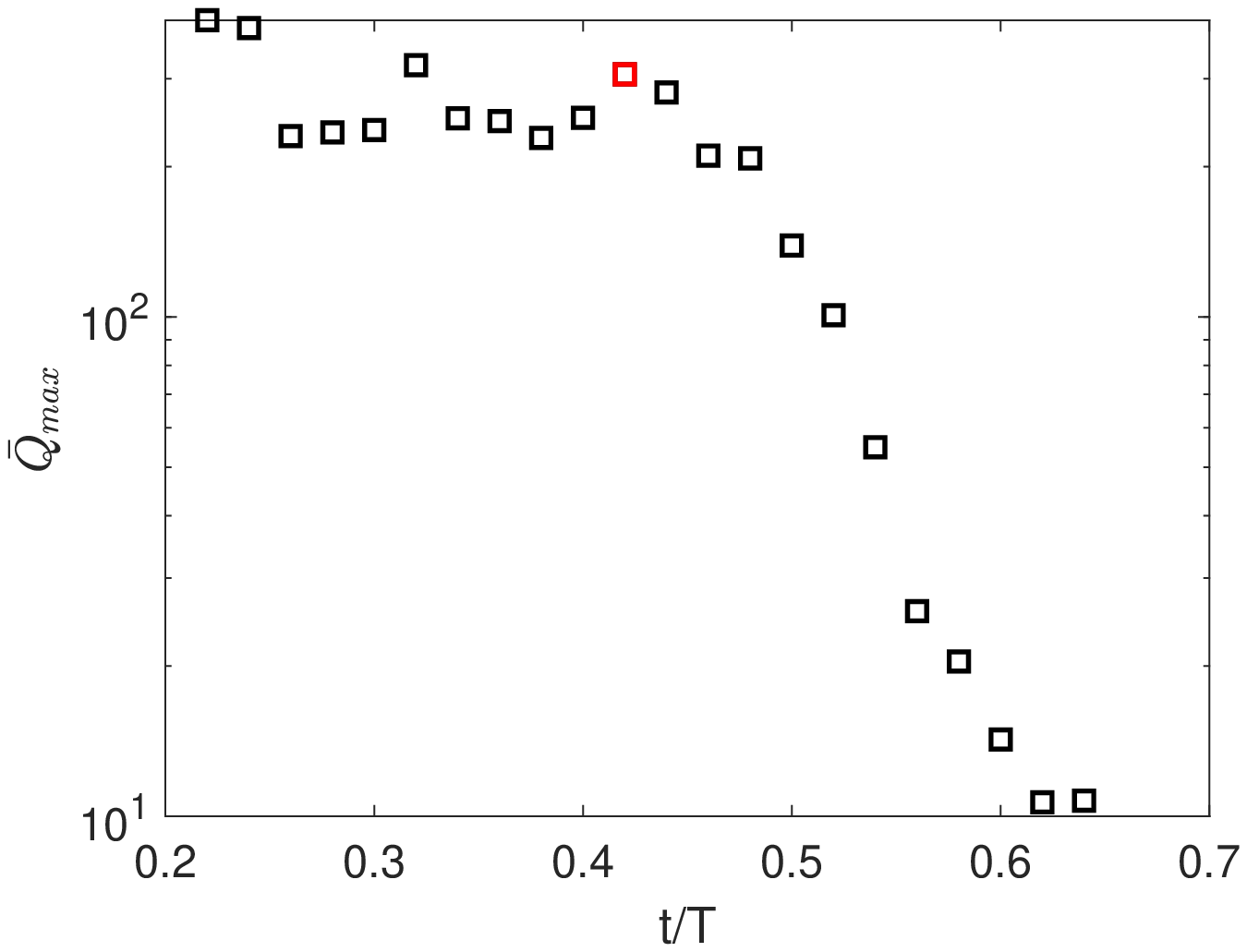}
\includegraphics[width=0.45\textwidth]{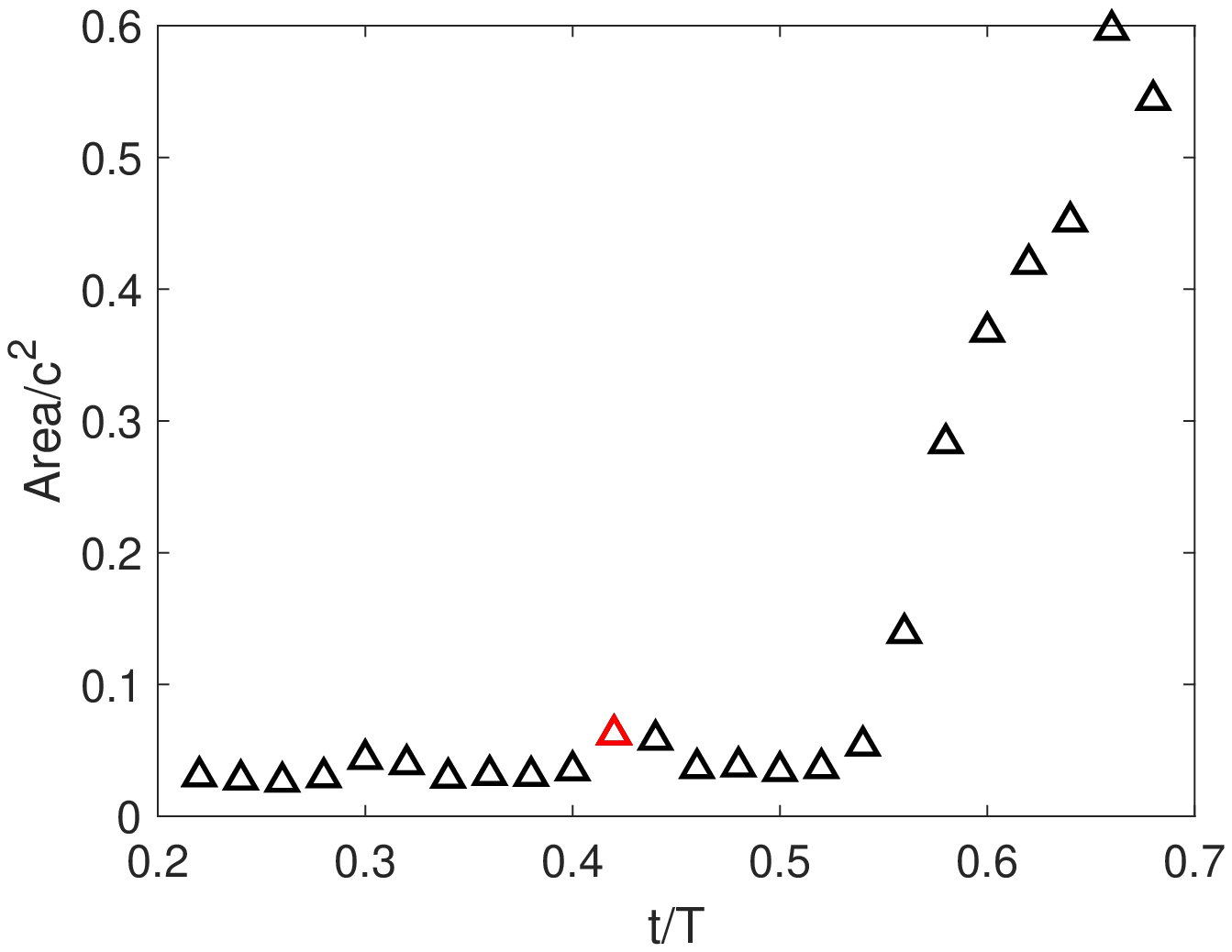}
\caption{An example of the evolution of the vortex strength, $\bar{Q}_{max}$ (left), and area (right), as a function of time. The red marker indicates the point of separation of the LEV from the leading edge, determined by a visual assessment of the vorticity plot. The foil's operating kinematics are: $f^* = 0.12, \theta_0 = 85^\circ, h_0/c = 1$, $\AT = 0.84$.}
\label{fig:maxQvsTime}
\end{figure}

\clearpage \begin{figure}
\centering
\includegraphics[width=0.5\textwidth]{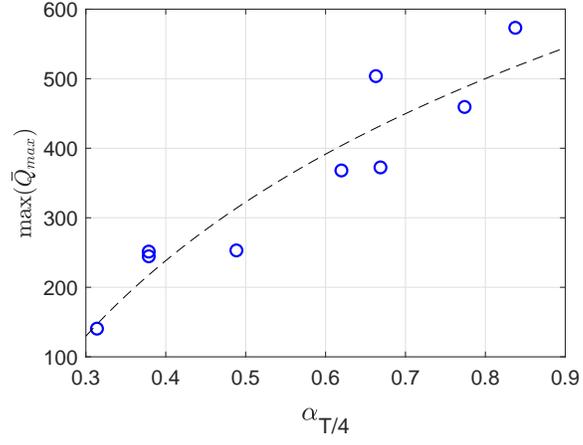}
\caption{Strength of the highest $\bar{Q}_{max}$ throughout the cycle, or $\max(\bar{Q}_{max})$, for various values of $\AT$. The best fit line follows the equation $377.51 \log(\AT) + 584.29$.}
\label{fig:maxQvsAlpha}
\end{figure}

\begin{figure*}
\centering
\subfloat[ ]{\label{fig:regime1Vort}\includegraphics[width=0.40\textwidth]{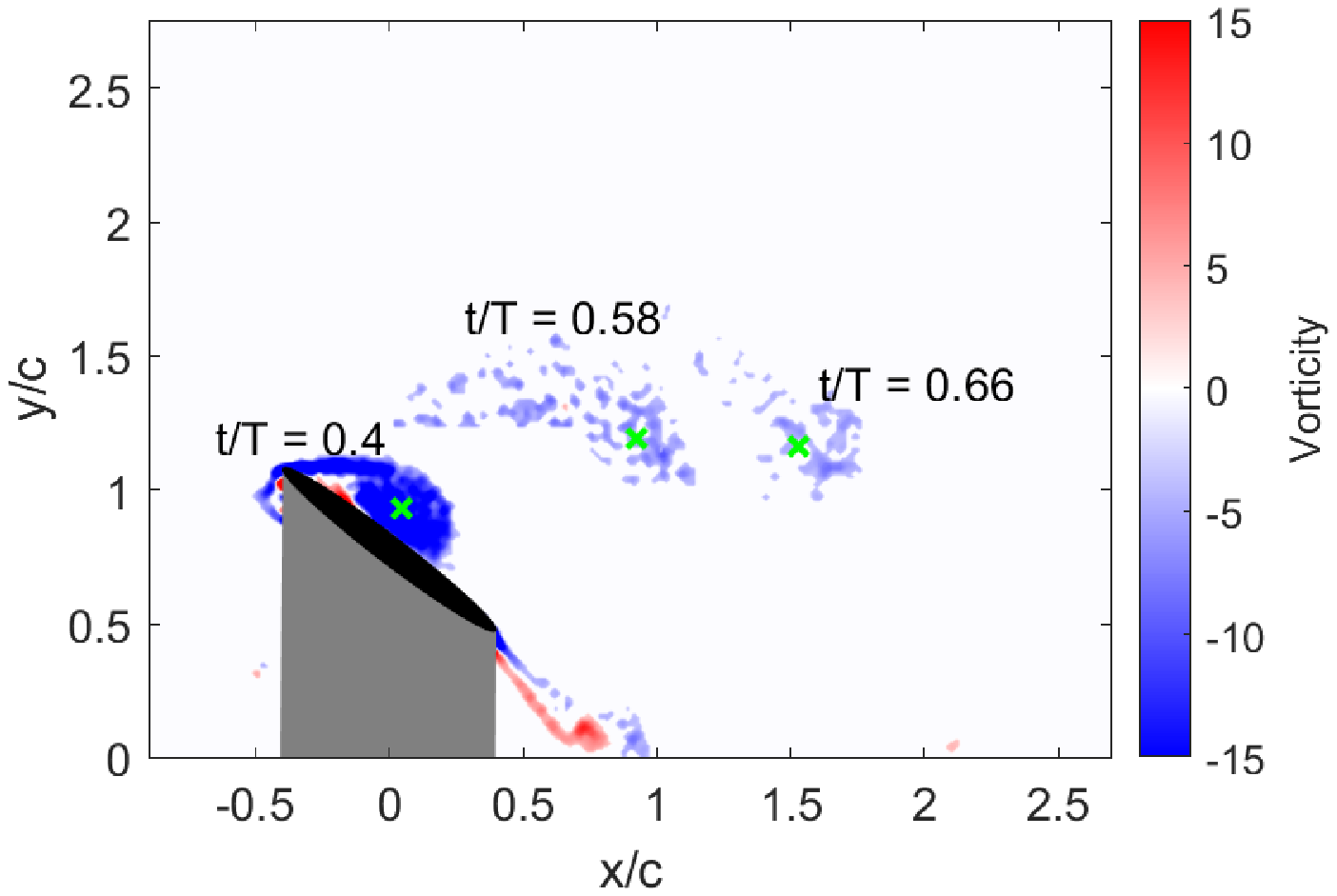}}
\subfloat[ ]{\label{fig:regime1Traj}\includegraphics[width=0.58\textwidth]{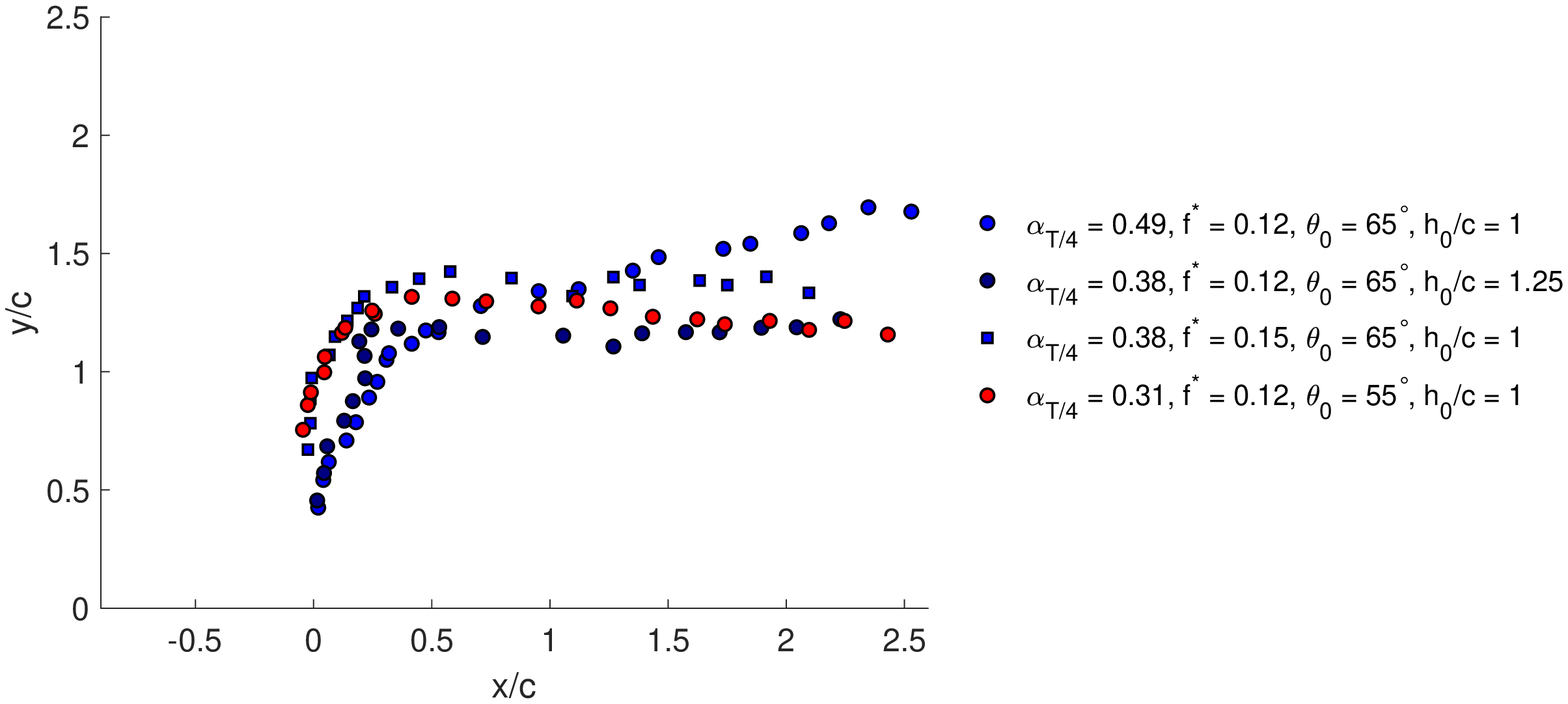}}\\
\subfloat[ ]{\label{fig:regime2Vort}\includegraphics[width=0.40\textwidth]{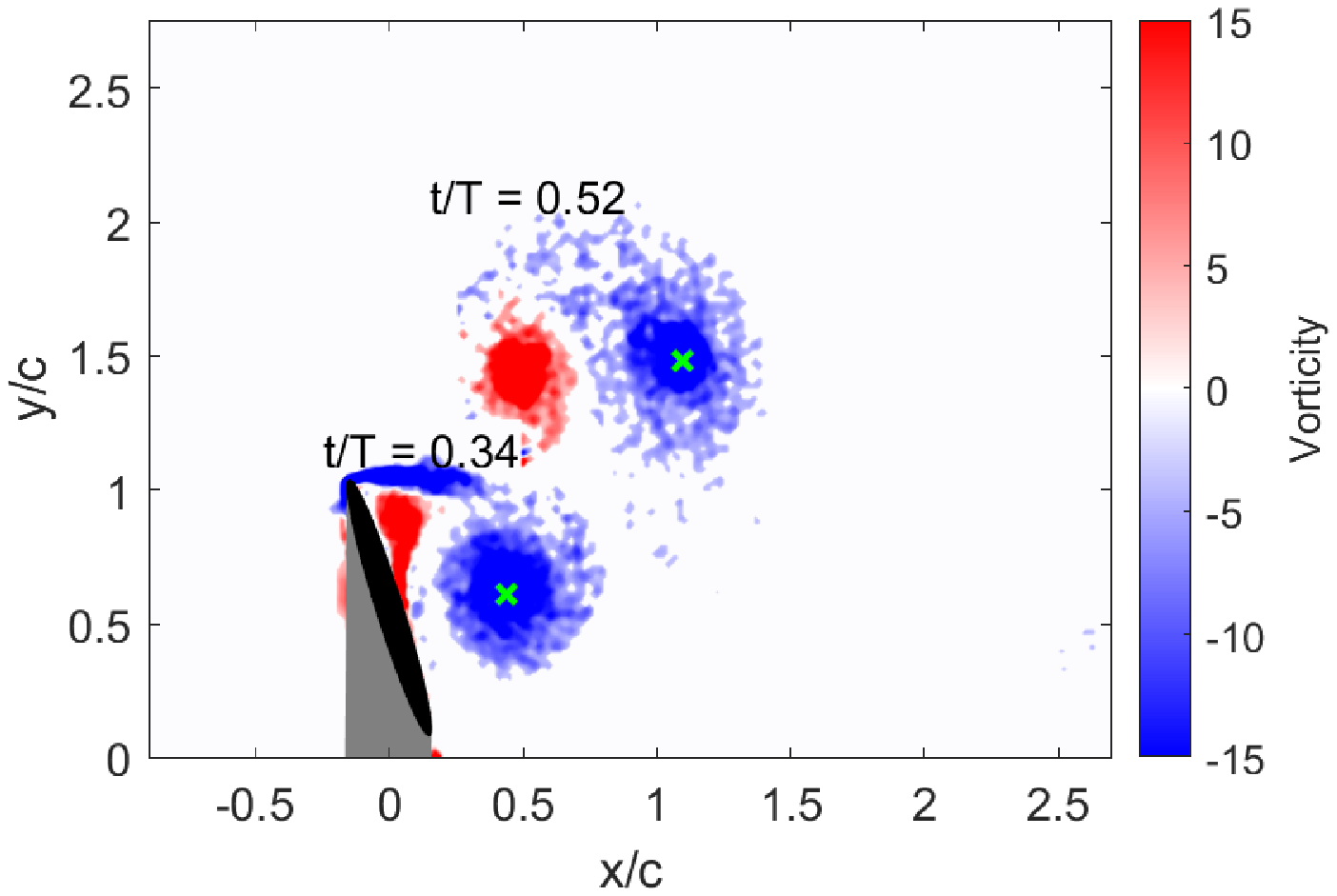}}
\subfloat[ ]{\label{fig:regime2Traj}\includegraphics[width=0.58\textwidth]{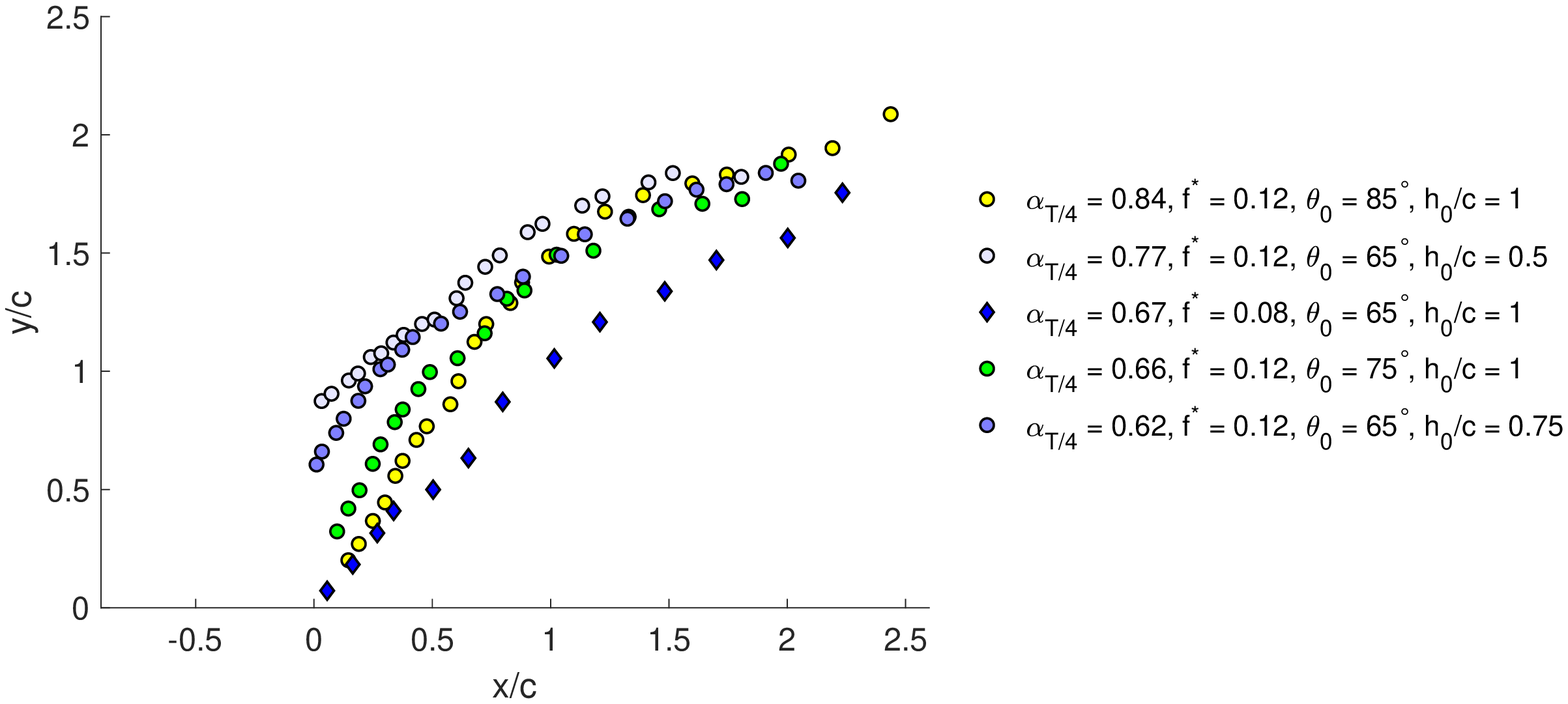}}
\caption{a) Snapshots at different points in time of a typical PIV vorticity plot of a LEV developing under the LEV regime. The tracked positions of the LEV centroid are indicated by green crosses. The foil kinematics for this example are: $f^* = 0.12$, $\theta_0 = 55^\circ$, $h_0/c = 1$. b) Trajectories of 4 different kinematics in LEV regime. The vortices follow a steep upwards motion early in the trajectory. After separation, LEVs convect downstream with relatively small y-displacements. c) Snapshots at different points in time of a typical PIV vorticity plot of a vortex pair developing under the LEV+TEV regime. Only the LEV is tracked. The foil kinematics for this example is: $f^* = 0.12$, $\theta_0 = 85^\circ$, $h_0/c = 1$. d) Trajectories of 5 different kinematics in the LEV+TEV regime. The initial steep upwards motion is still seen. After separation, however, LEVs continue to retain positive y-velocity.}
\label{fig:regimes}
\end{figure*}

\clearpage \begin{figure}
\centering
\includegraphics[width=0.6\textwidth]{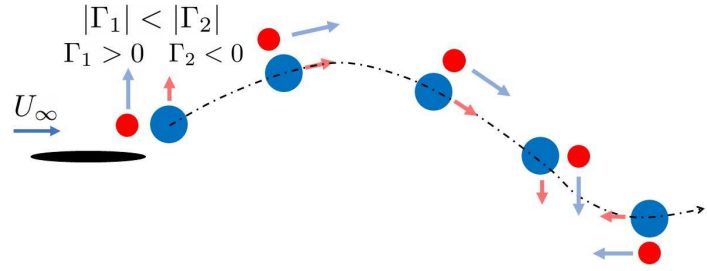}
\caption{The interaction between the LEV and the TEV and the resultant trajectory characteristic to LEV+TEV regime. The circles represent the two vortices with the colors indicating the sign of circulation. The arrow indicates the direction of induced velocity.}
\label{fig:summaryLEVTEVRegime}
\end{figure}

\clearpage \begin{figure}
\centering
\includegraphics[width=0.6\textwidth]{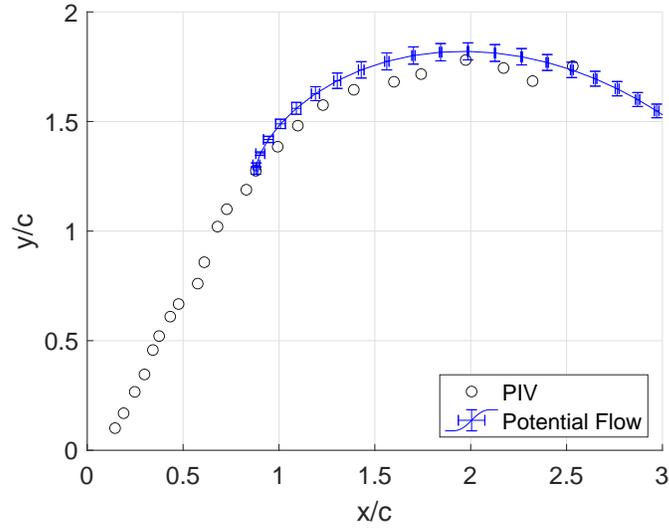}
\caption{An example of the comparison between a trajectory predicted from the potential flow model and a trajectory obtained experimentally. The error bars represent the standard deviation of Monte Carlo simulation of the potential flow model. The kinematics of the foil a) $f^* = 0.12$, $\theta_0 = 85^\circ$, $h_0/c = 1$. }
\label{fig:experimentalPotential}
\end{figure}

\clearpage \begin{figure}
\subfloat[]{\label{fig:PotentialA}\includegraphics[scale=.4]{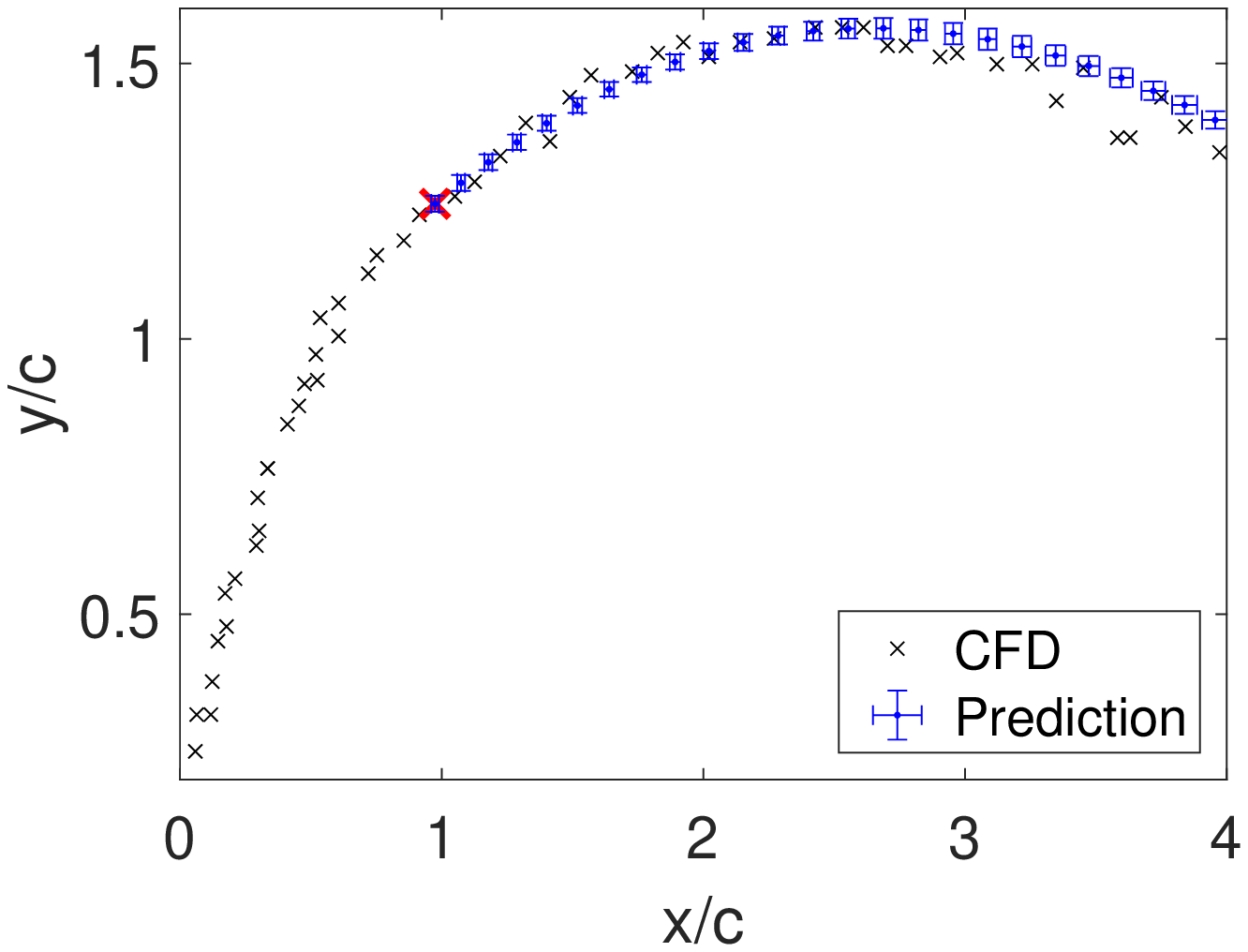}}
\subfloat[]{\label{fig:PotentialB}\includegraphics[scale=.4]{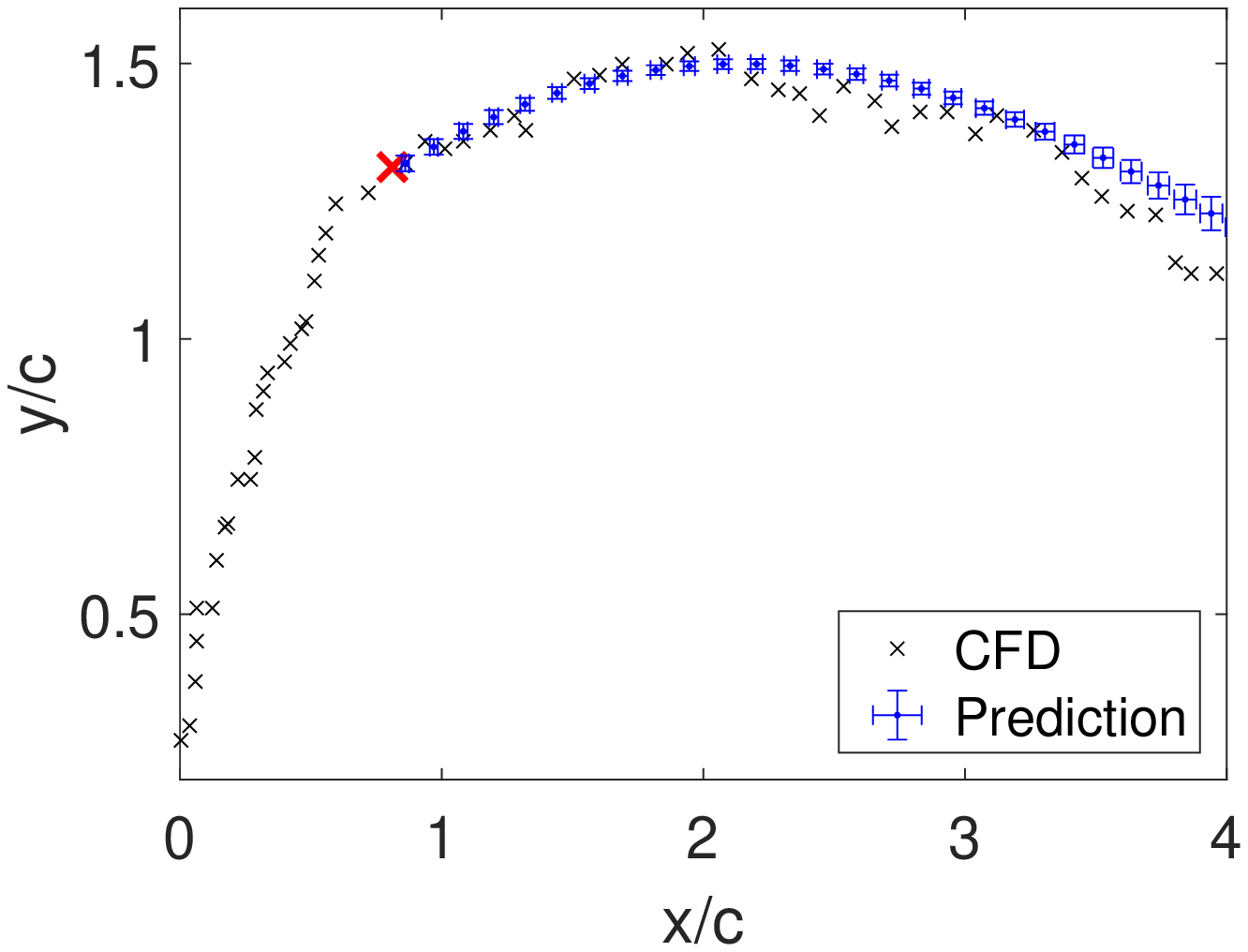}}
\\
\subfloat[]{\label{fig:PotentialC}\includegraphics[scale=.4]{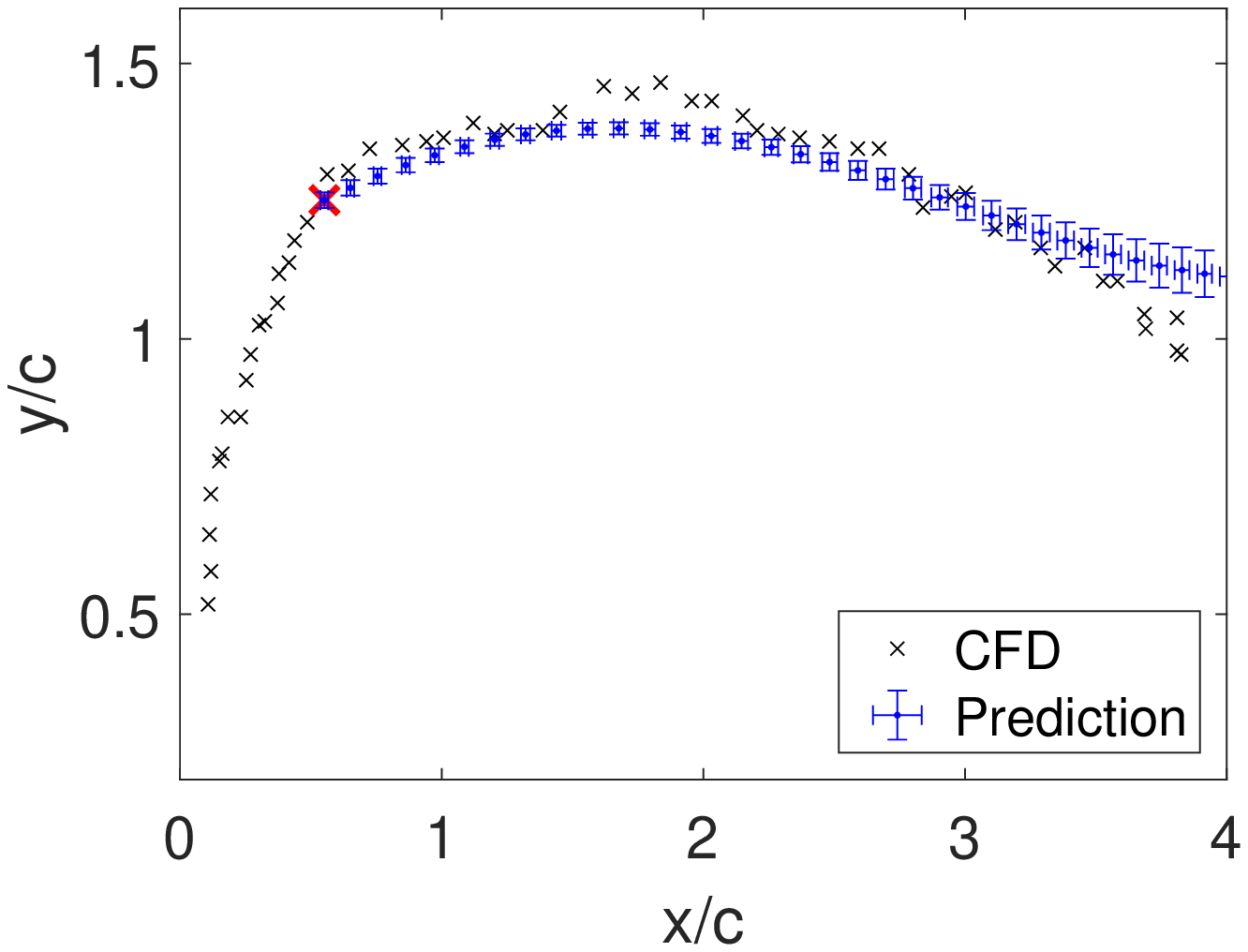}}
\subfloat[]{\label{fig:PotentialD}\includegraphics[scale=.4]{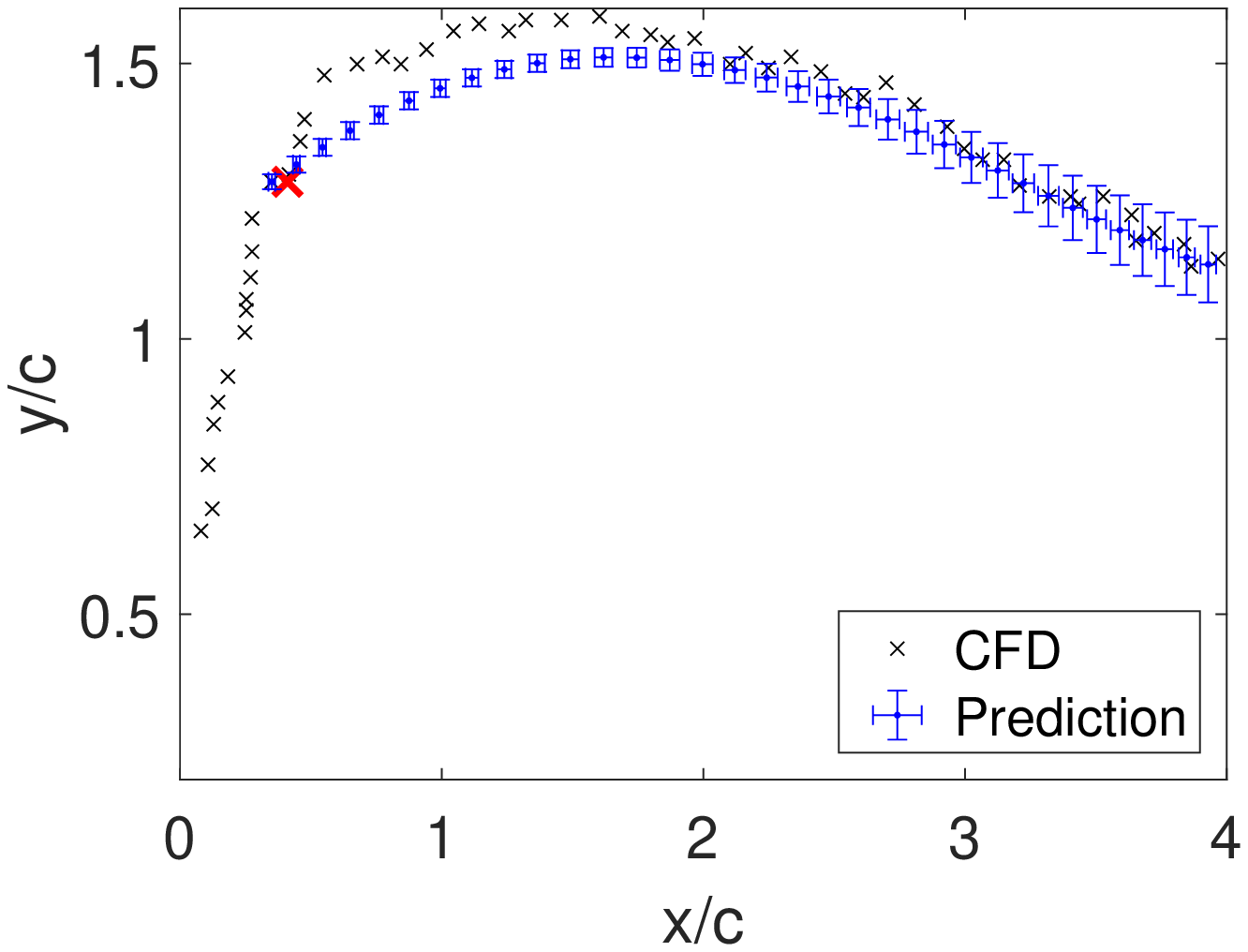}}
\caption{Four examples of the comparison between a trajectory predicted from the potential flow model and a trajectory obtained from the vortex formations in the CFD data. The error bars represent the standard deviation of Monte Carlo simulation. Red cross indicates the point of separation of the LEV from the leading edge. The kinematics of the foil are: a) $f^* = 0.10$, $\theta_0 = 65^\circ$, $h_0/c = 1$. b) $f^* = 0.11$, $\theta_0 = 65^\circ$, $h_0/c = 1$. c) $f^* = 0.12$, $\theta_0 = 65^\circ$, $h_0/c = 1$. d) $f^* = 0.15$, $\theta_0 = 75^\circ$, $h_0/c = 1$.}
\label{fig:PotentialFlowModel}
\end{figure}

\clearpage \begin{figure}
\centering
\subfloat[x-coordinate]{\label{fig:xPrediction}\includegraphics[scale=.4]{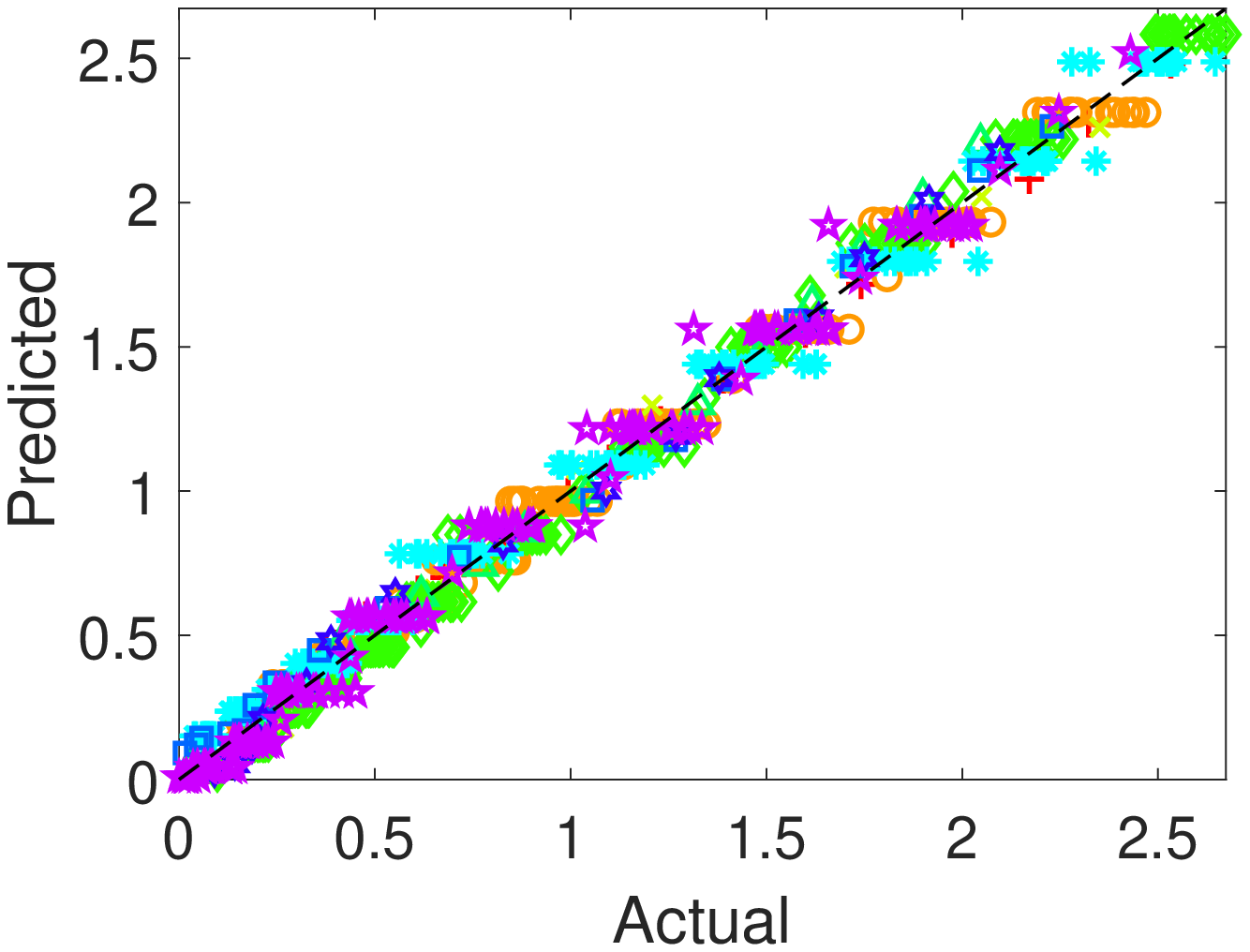}}
\subfloat[y-coordinate]{\label{fig:yPrediction}\includegraphics[scale=.4]{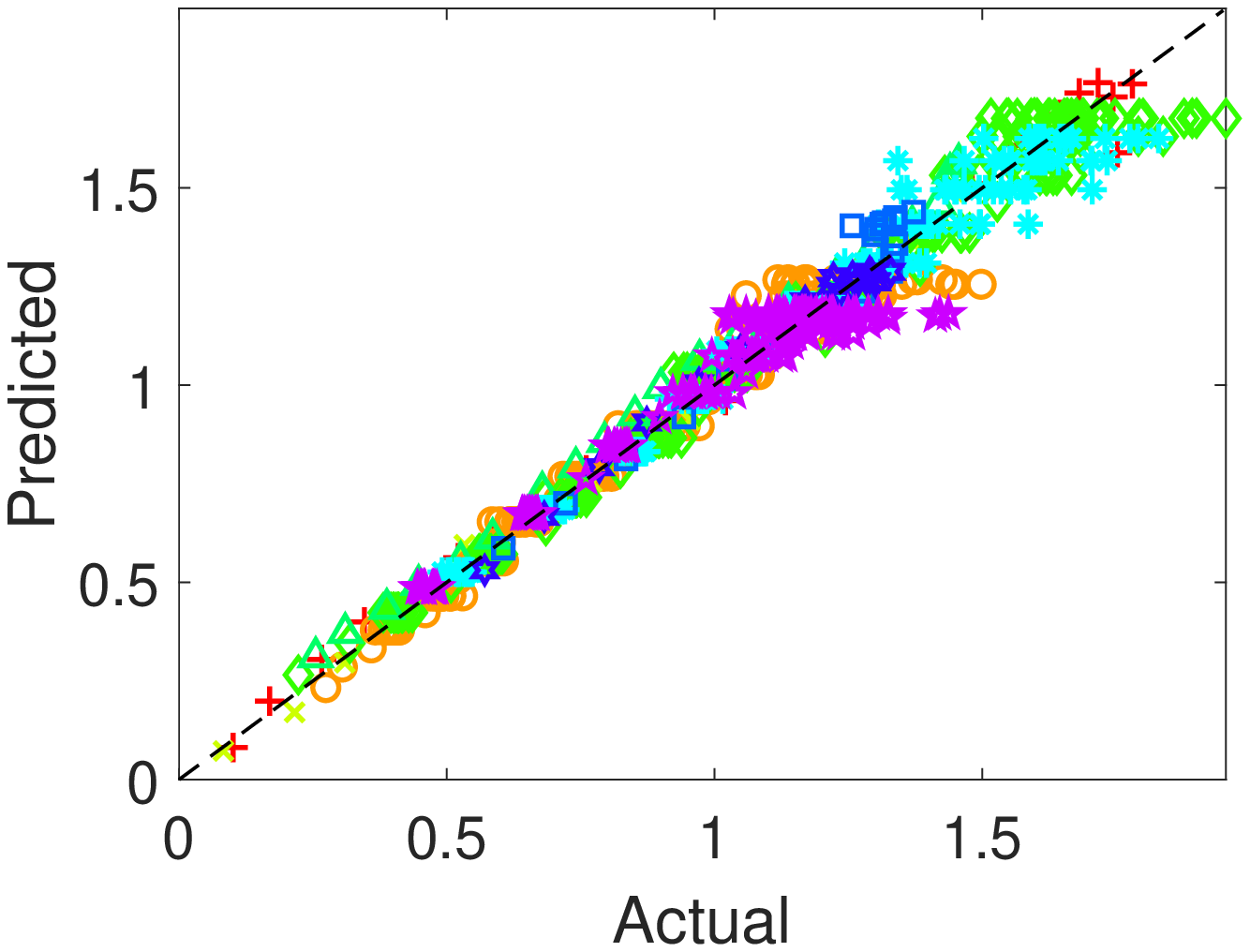}}\\
\subfloat[Q/max($\bar{Q}_{max}$)]{\label{fig:qPrediction}\includegraphics[scale=.4]{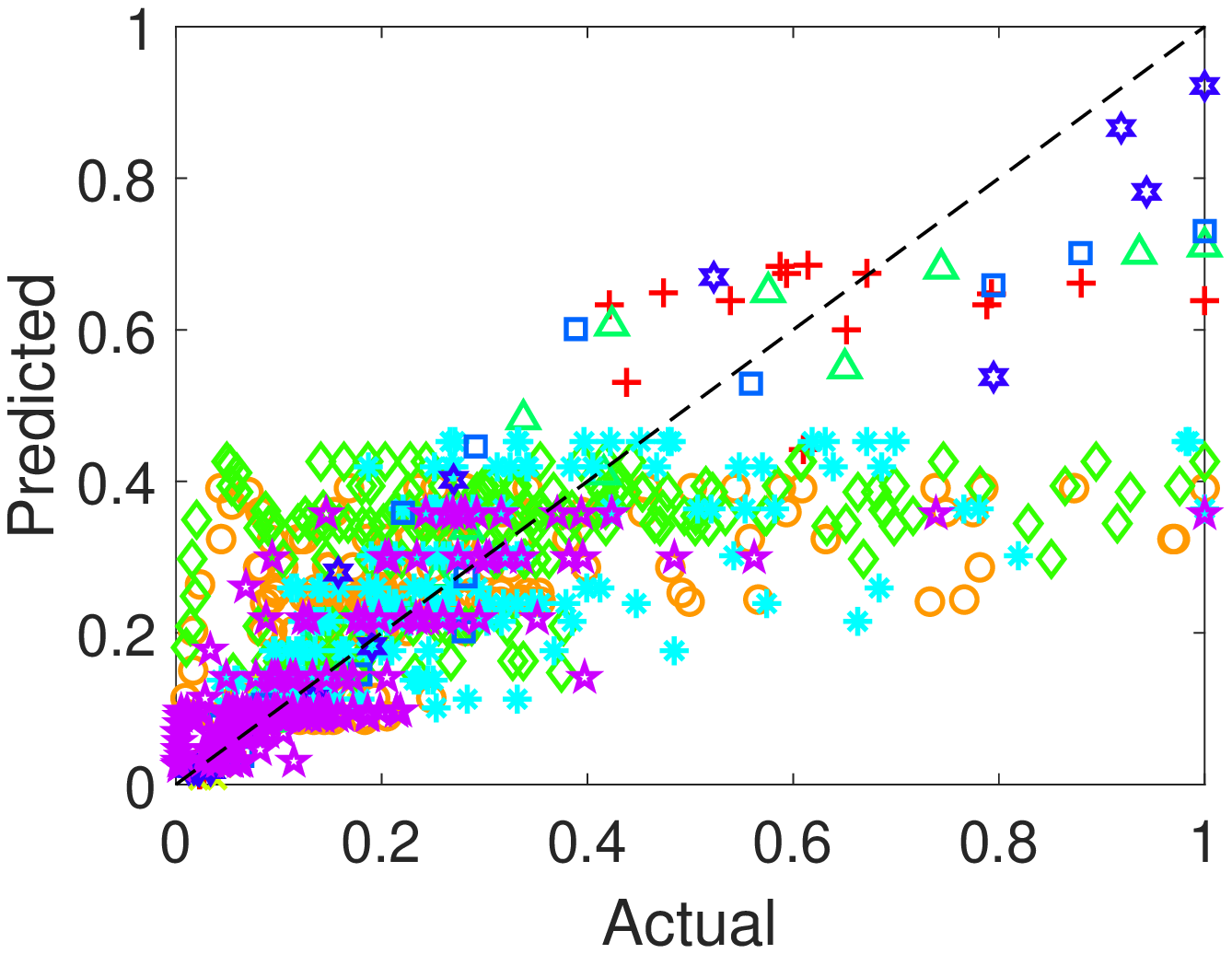}}
\subfloat[Ellipse fit - major axis]{\label{fig:axisAPrediction}\includegraphics[scale=.4]{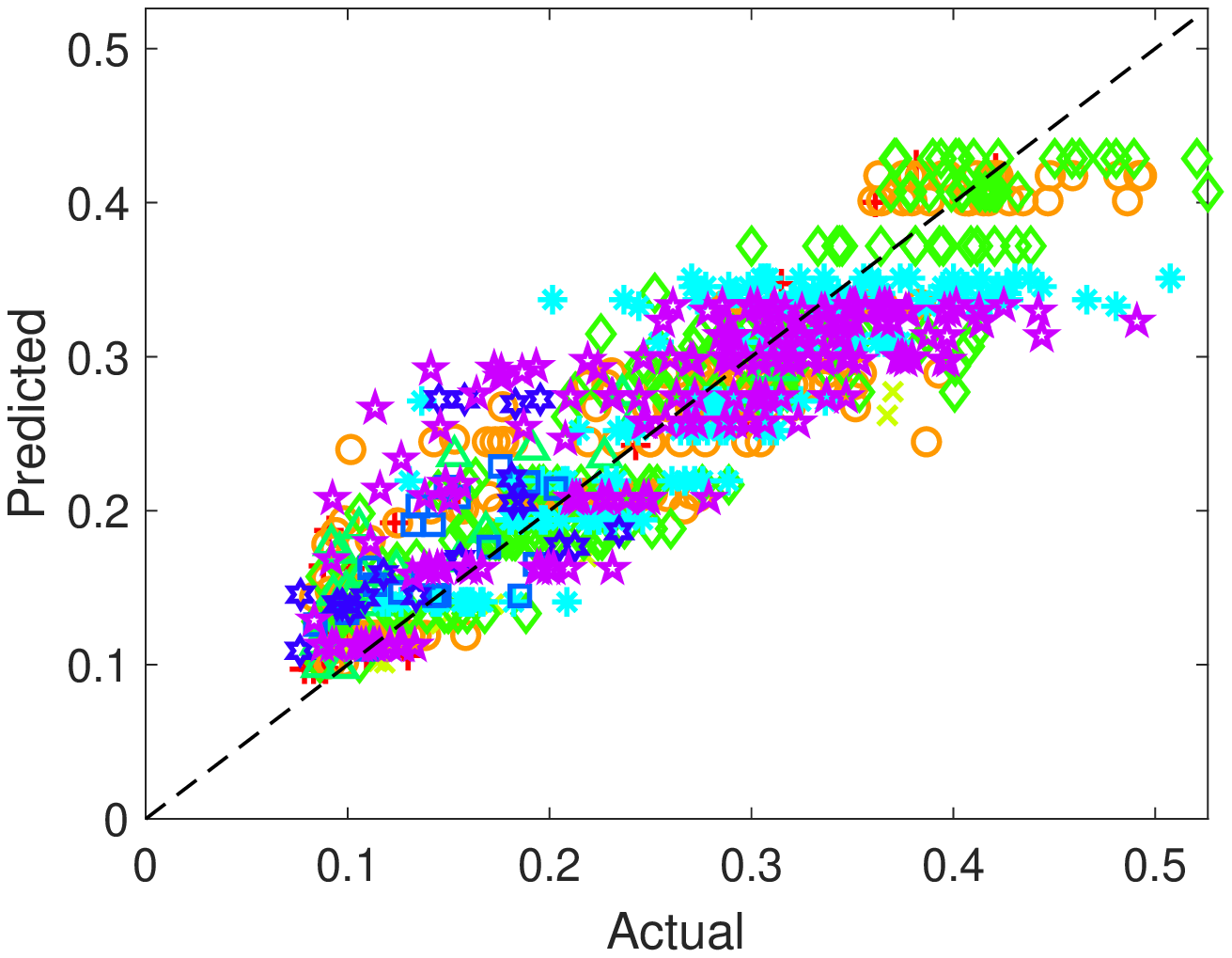}}\\
\subfloat[Ellipse fit - minor axis]{\label{fig:axisBPrediction}\includegraphics[scale=.4]{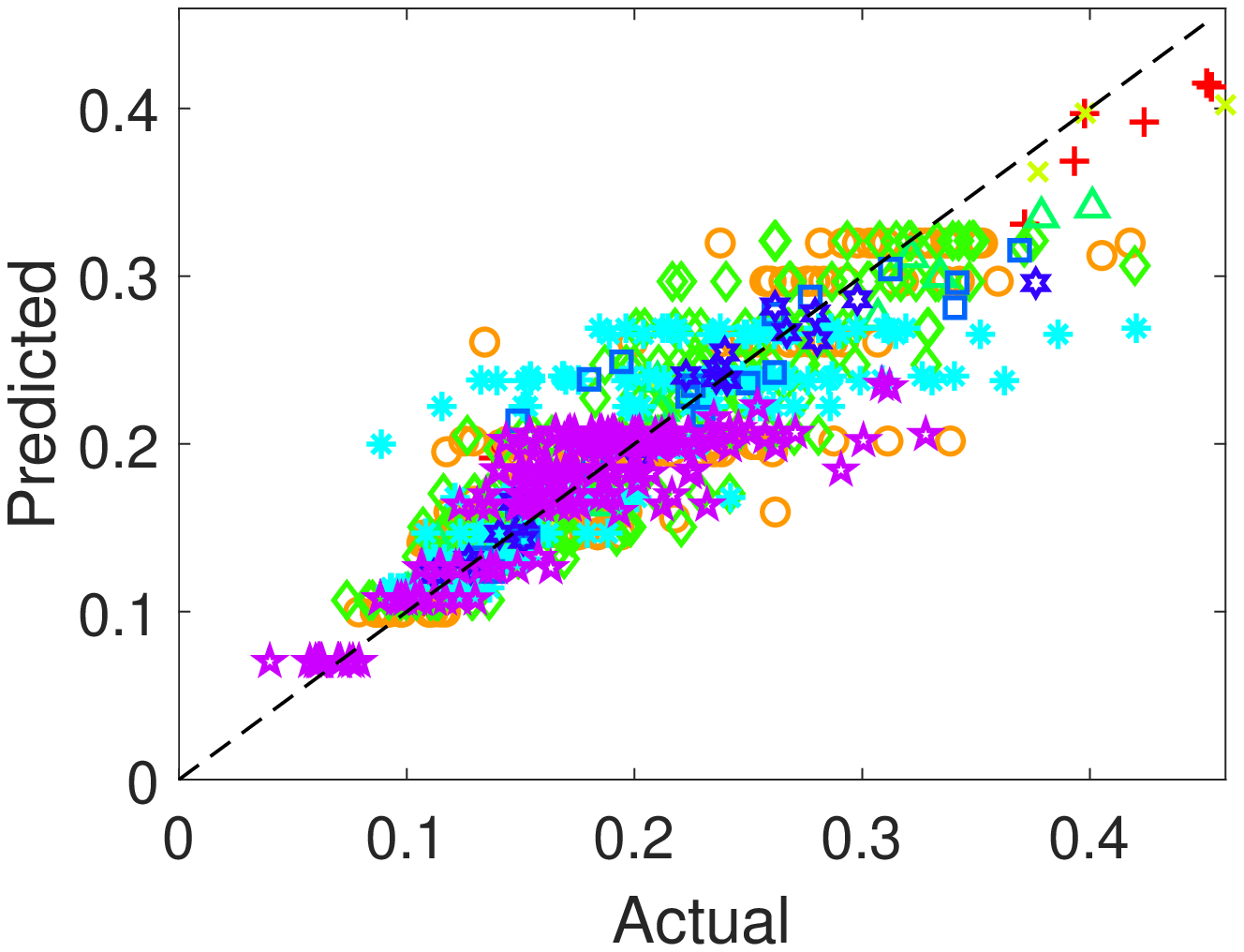}}
\subfloat{\label{fig:legend}\includegraphics[scale=.4]{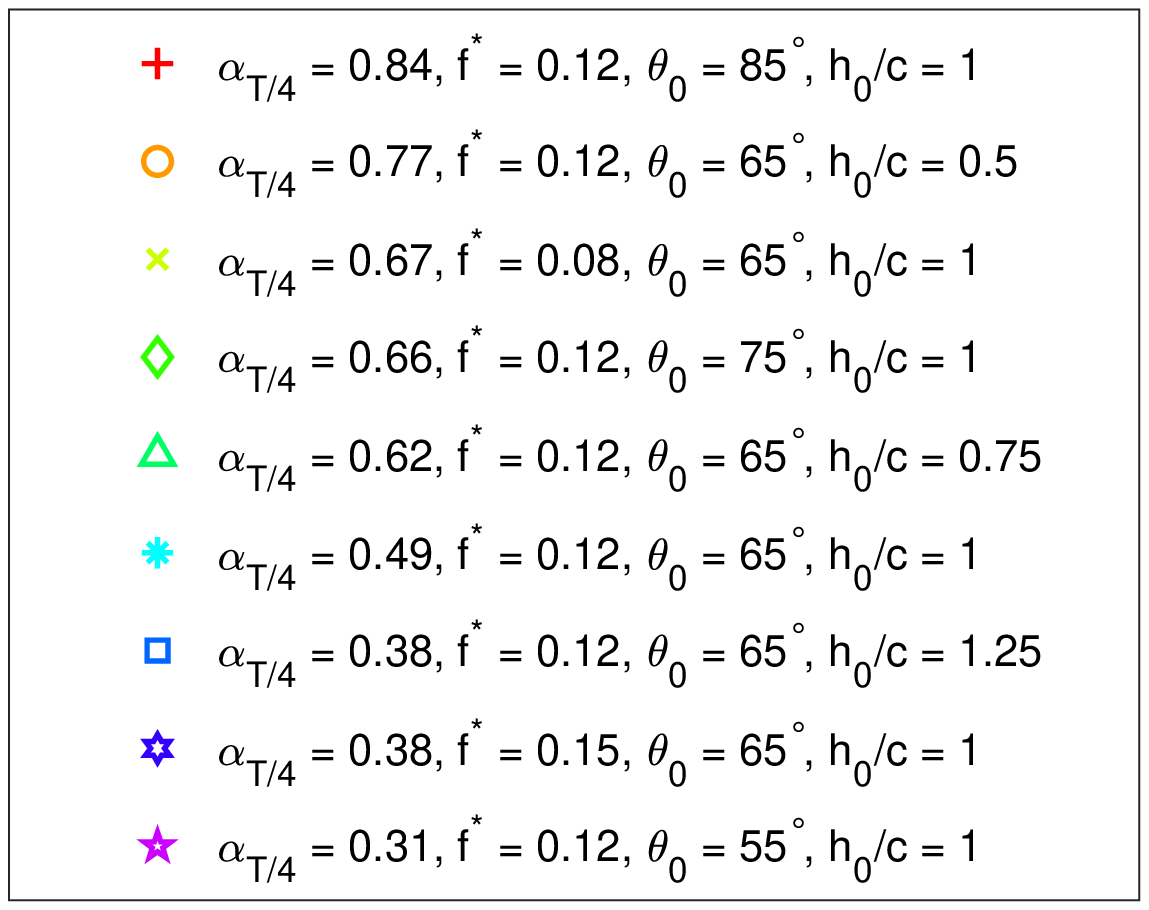}}\\

\caption{The machine learning performance per predicted variable as demonstrated by the ``Predicted vs Actual'' plot. The black diagonal line represents a perfect prediction, while the plot markers represent the actual true response of the machine learning algorithm for each PIV frame used in the training. Due to the non-phase-averaged nature of the training data, multiple actual values exist for each kinematics resulting in a horizontal ``line" in the plots. The length of this line represents the range of the actual value within the training data. The predicted values are: a) centroid x-coordinate, b) centroid y-coordinate, c) $Q/\max(\bar{Q}_{max})$, d) major axis of equivalent ellipse, e) minor axis of equivalent ellipse.}
\label{fig:PredictedvsActual}
\end{figure}

\clearpage \begin{figure}
\centering
\centering 
 \subfloat[Measured]{\includegraphics[width=0.4\textwidth]{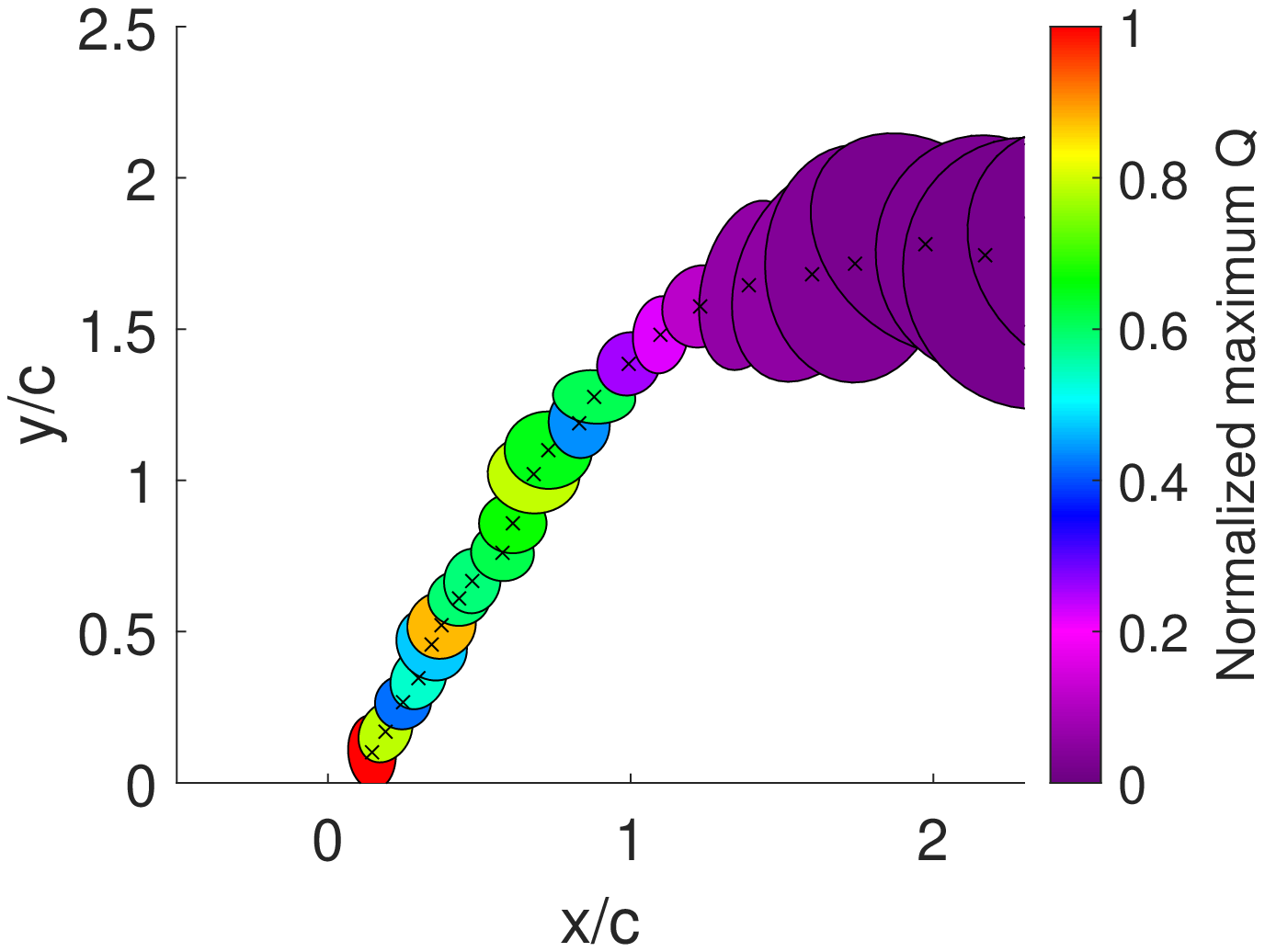}\label{fig:PIVCheck}}
 \subfloat[Predicted]{\includegraphics[width=0.4\textwidth]{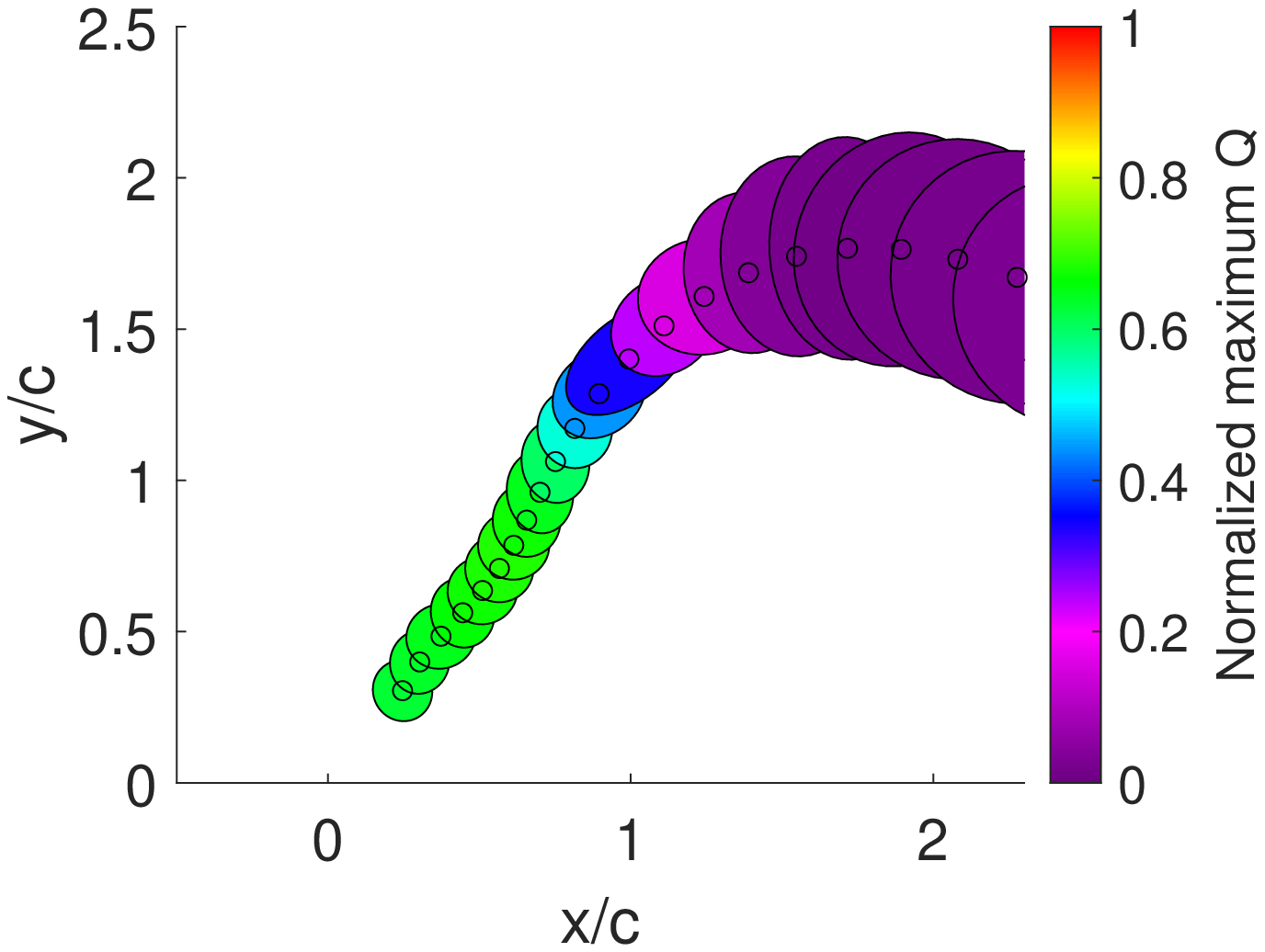}\label{fig:MLCheck}}
\caption{A comparison between the measured LEV trajectory (a) and the equivalent predicted trajectory (b) at foil kinematics $f^* = 0.12, \theta_0 = 85^\circ, h_0/c = 1$. Dark crosses and circles represent the centroids of the LEV. The ellipse represents the LEV's equivalent moment ellipses while their colors represent the maximum value of $Q$ that occurs within the vortex. The maximum $Q$-values are normalized by the highest value that occurs in the cycle.}
\label{fig:trainingExample}
\end{figure}

\clearpage \begin{figure}

\subfloat[Measured, Case i]{\label{fig:a}\includegraphics[scale=.34]{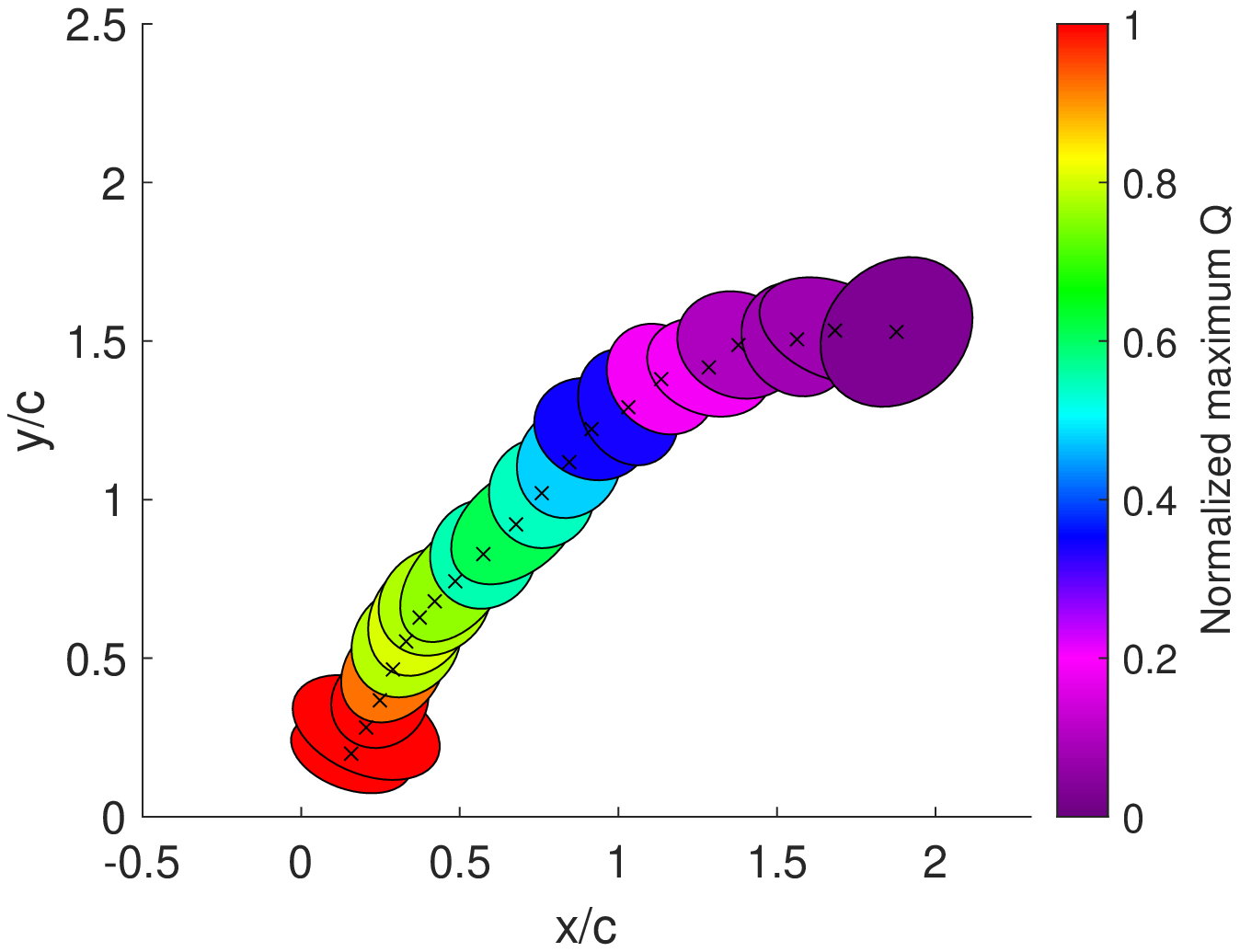}}
\subfloat[Predicted, Case i]{\label{fig:b}\includegraphics[scale=.34]{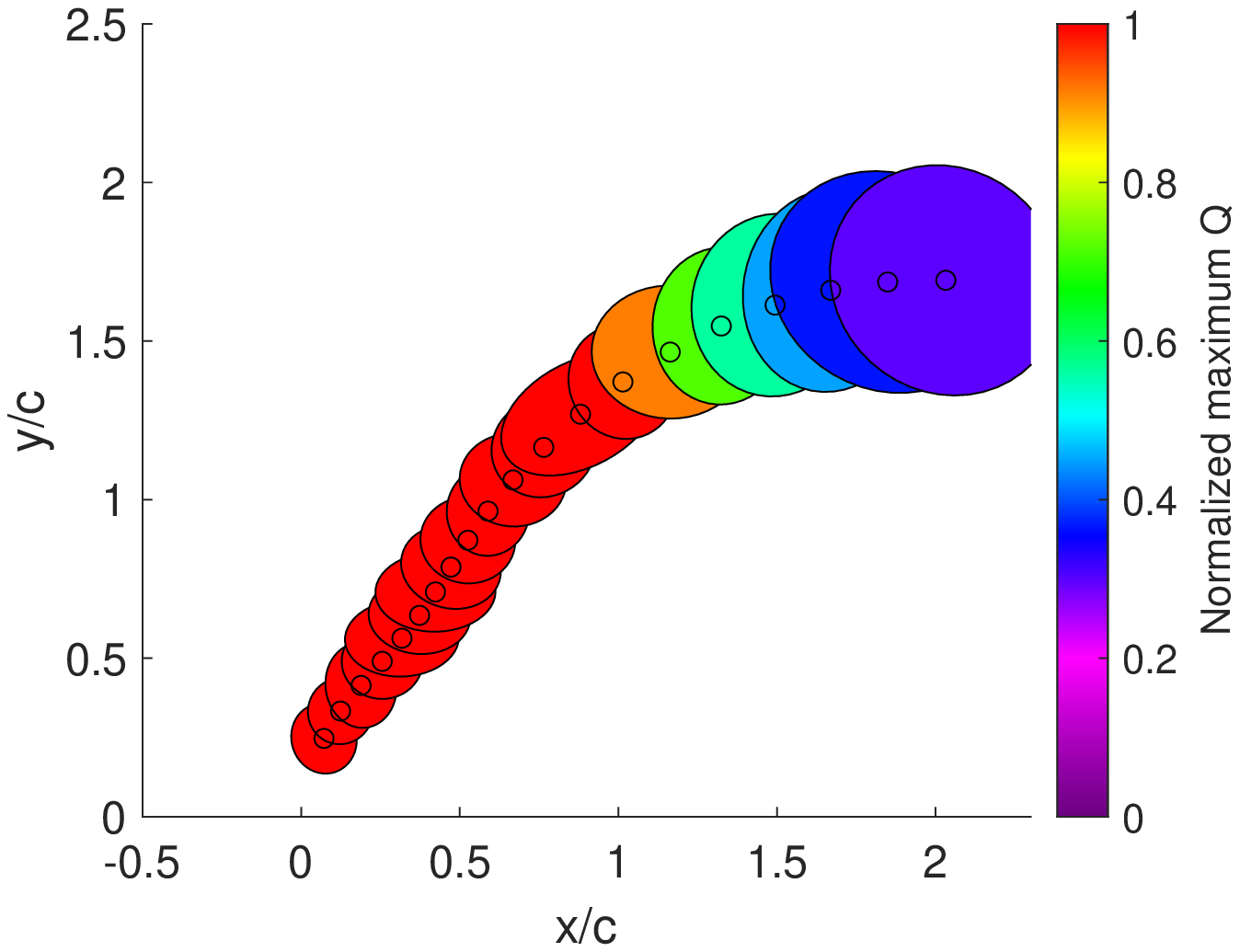}}\\
\subfloat[Measured, Case ii]{\label{fig:c}\includegraphics[scale=.34]{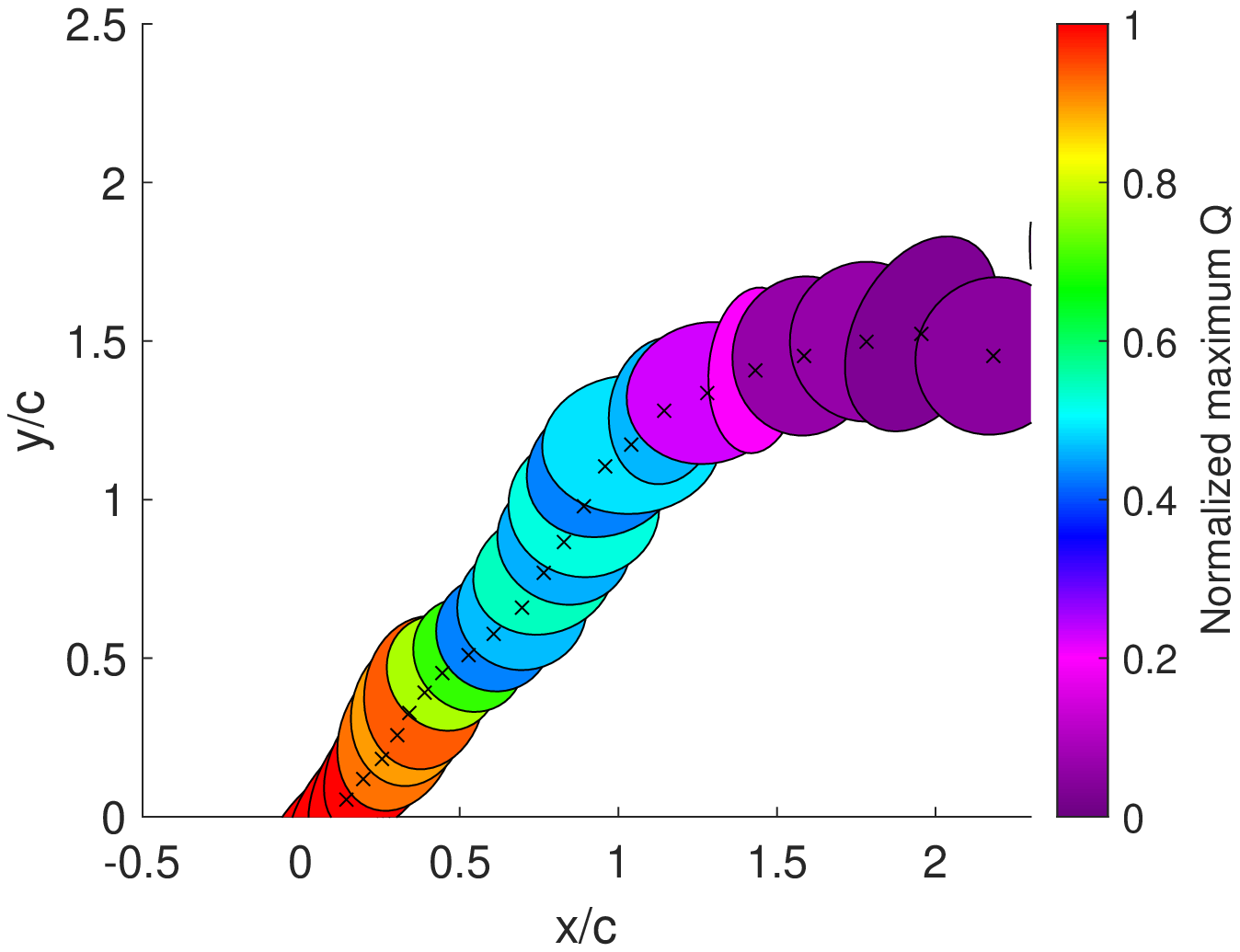}}
\subfloat[Predicted, Case ii]{\label{fig:d}\includegraphics[scale=.34]{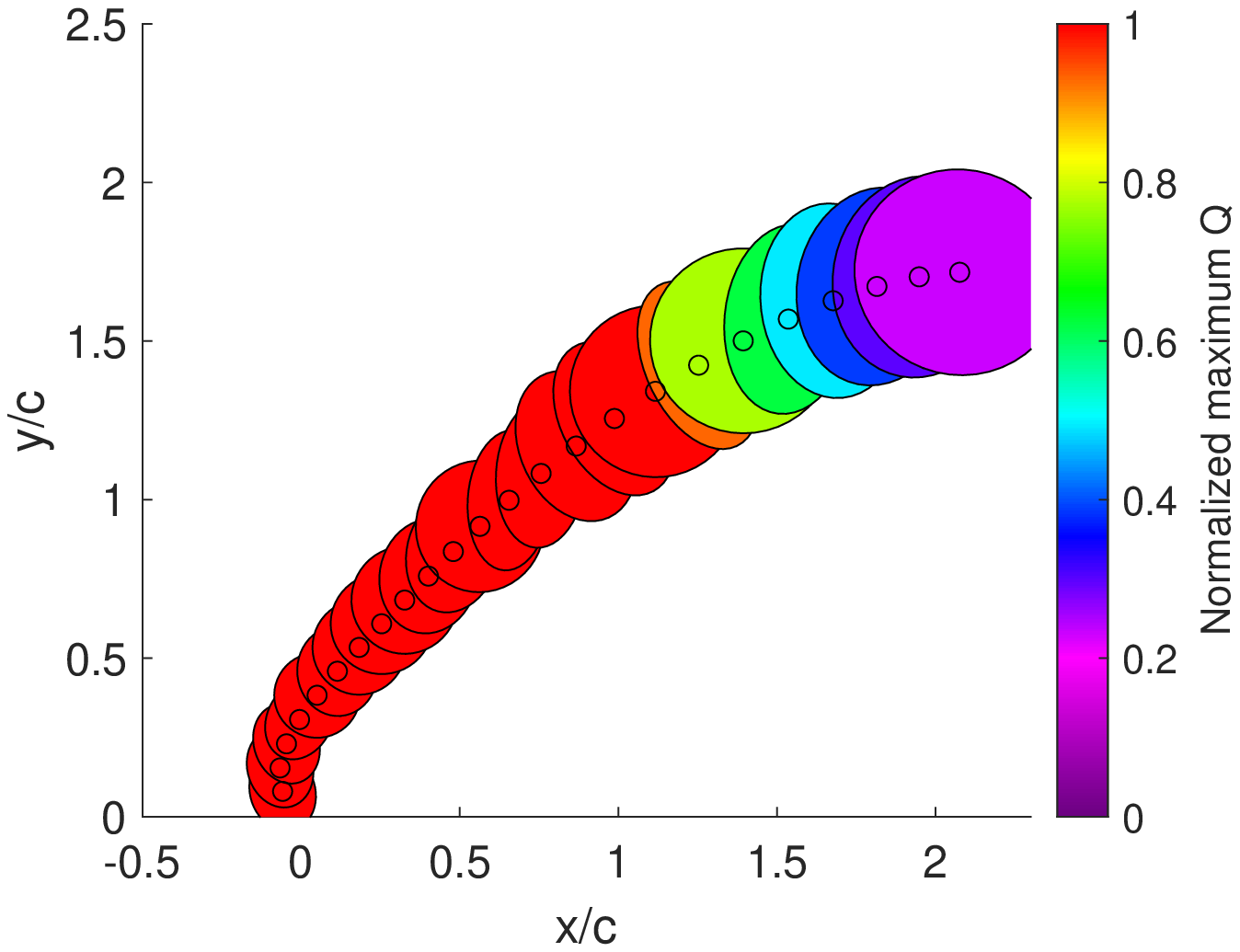}}\\
\subfloat[Measured, Case iii]{\label{fig:e}\includegraphics[scale=.34]{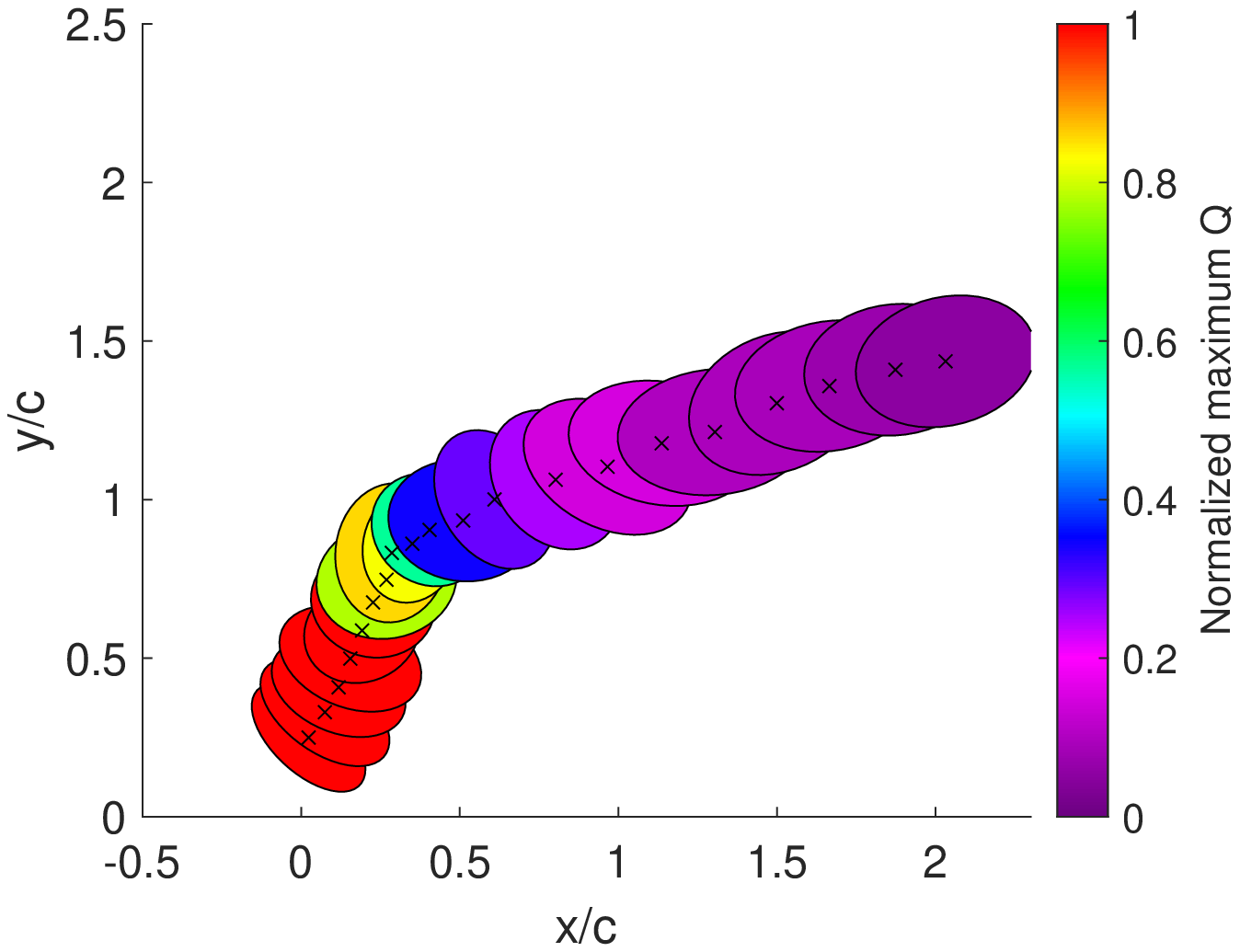}}
\subfloat[Predicted, Case iii]{\label{fig:f}\includegraphics[scale=.34]{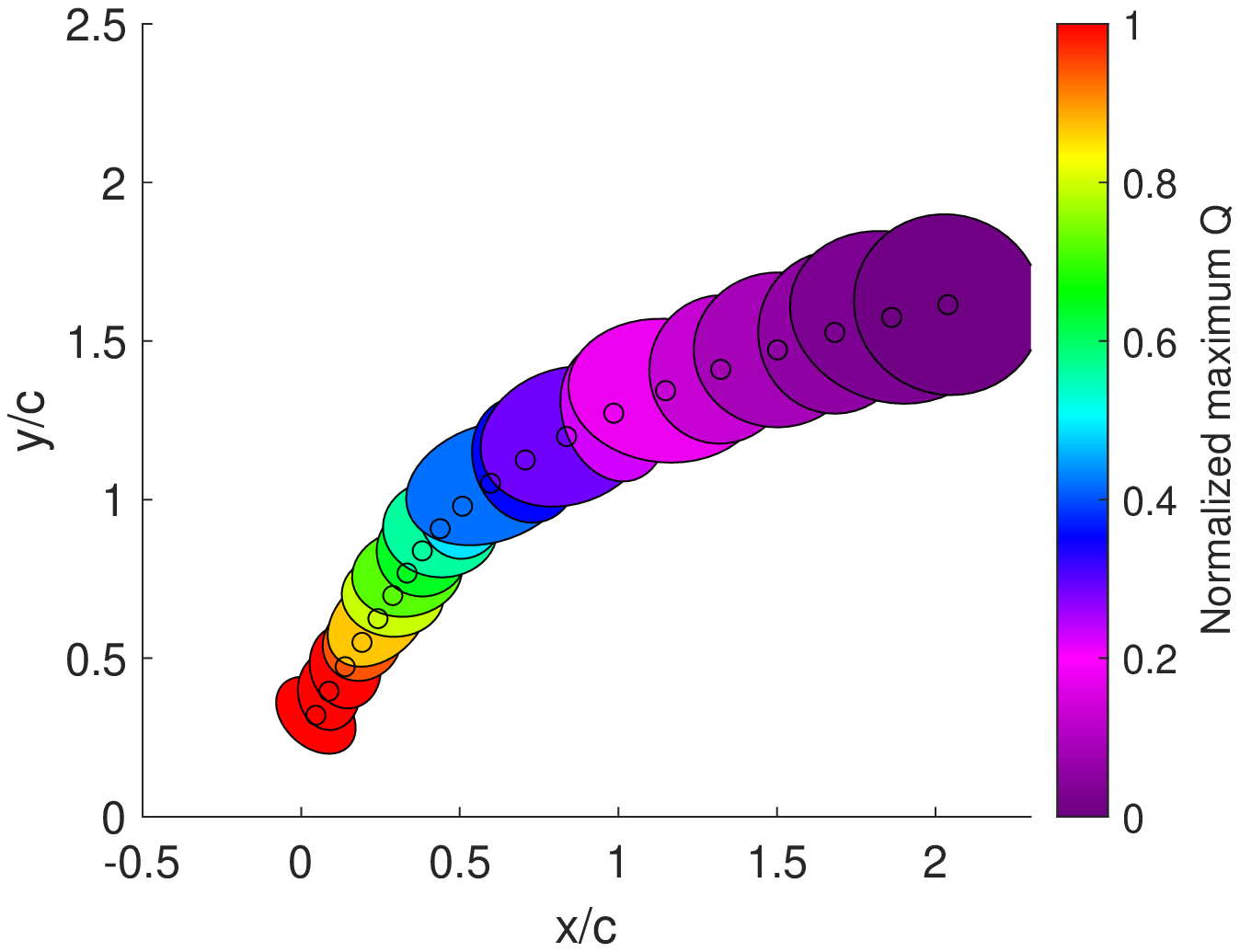}}\\
\subfloat[Measured, Case iv]{\label{fig:g}\includegraphics[scale=.34]{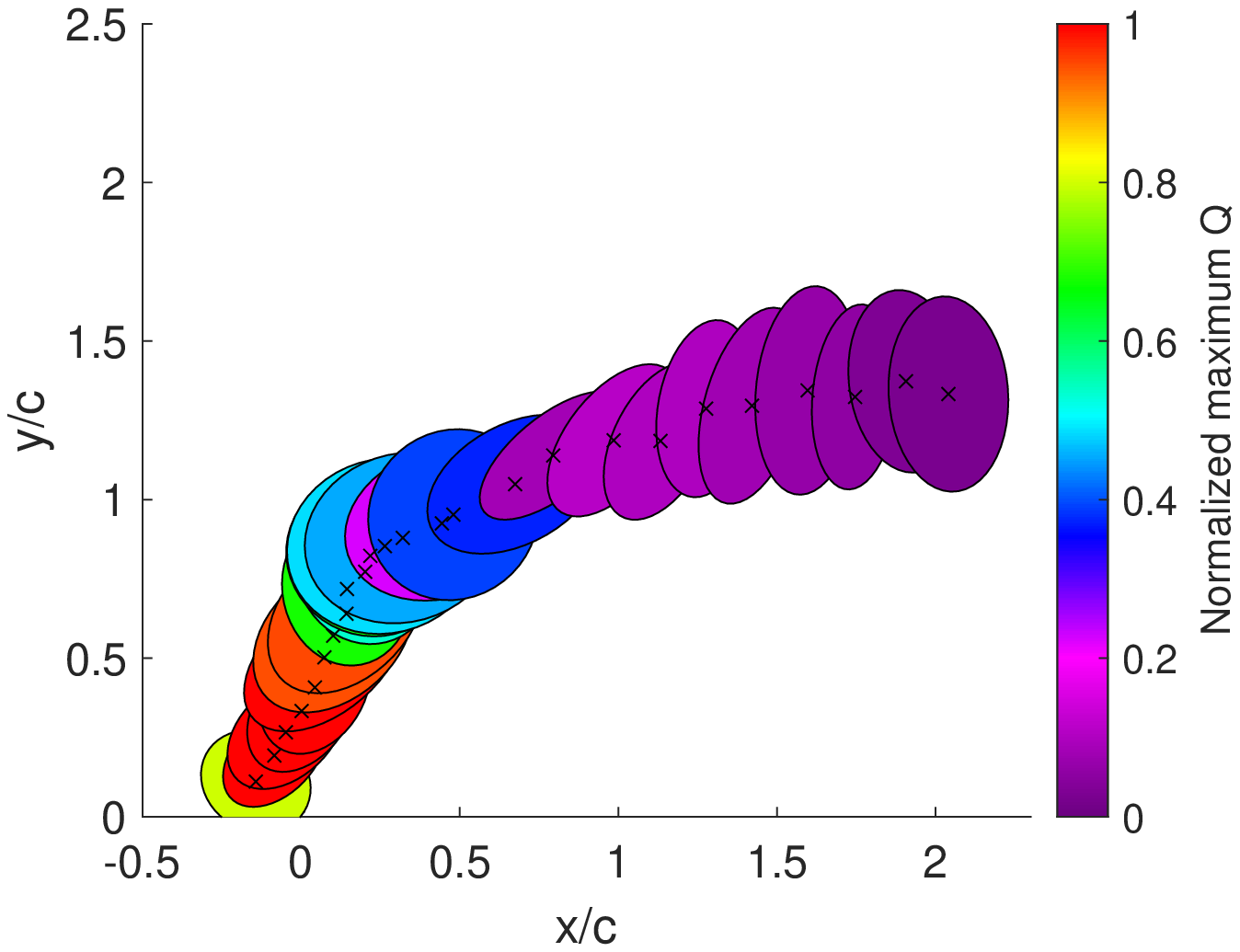}}
\subfloat[Predicted, Case iv]{\label{fig:h}\includegraphics[scale=.34]{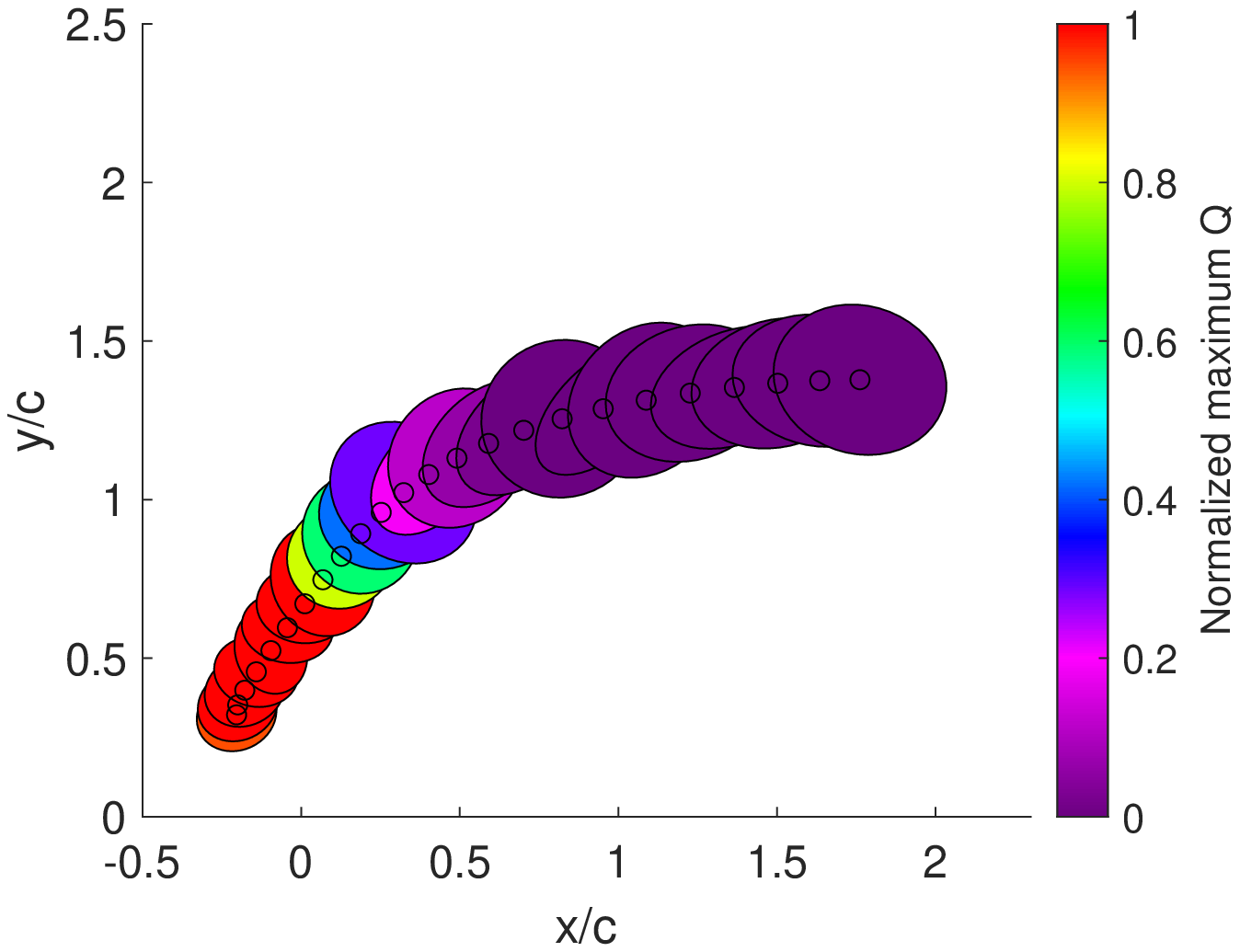}}
\caption{Comparison between predicted and actual LEV trajectory. Left: PIV data;  right: ML prediction. Dark crosses/circles represent the centroids of LEVs for PIV and ML, respectively. The color of the equivalent moment ellipse is based on the maximum $Q$-value in the field, normalized by the maximum $Q$-value over the entire cycle. Refer to Table \ref{table:verificationParameters} for foil kinematics.}
\label{fig:MLshowcase}
\end{figure}

\clearpage \begin{figure}
\subfloat[Case i]{\label{main:a}\includegraphics[scale=.4]{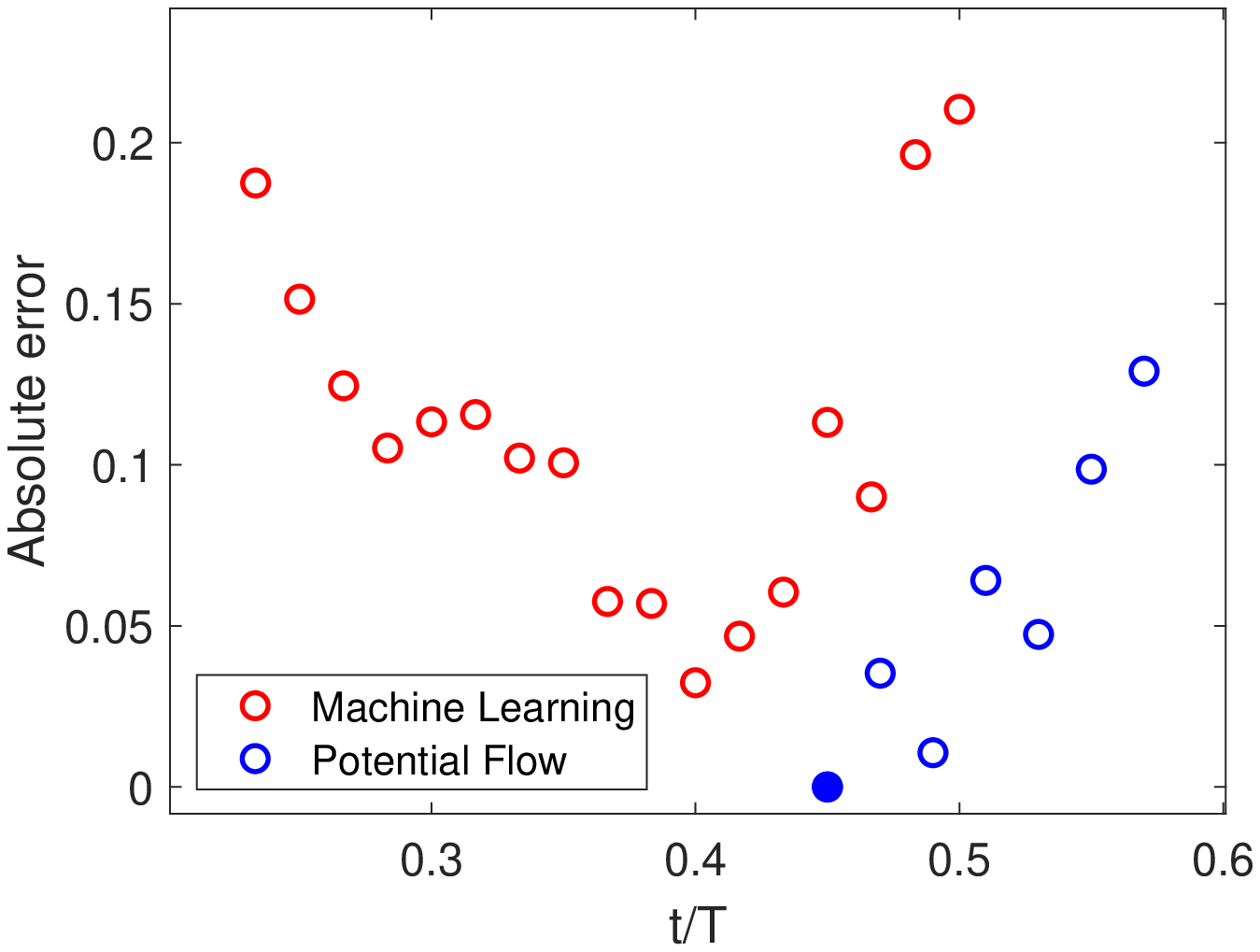}}
\subfloat[Case ii]{\label{main:b}\includegraphics[scale=.4]{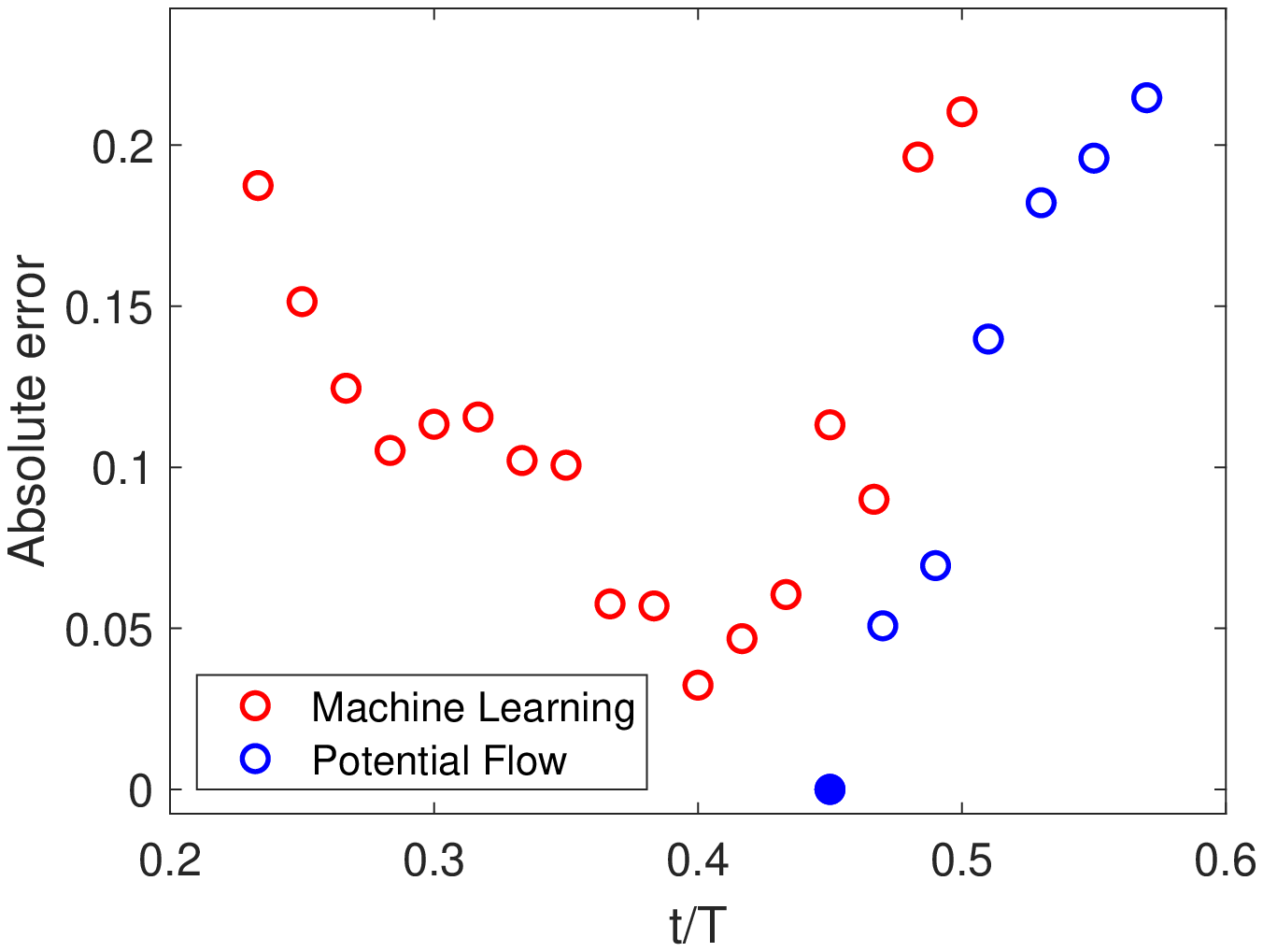}}\\
\subfloat[Case iii]{\label{main:c}\includegraphics[scale=.4]{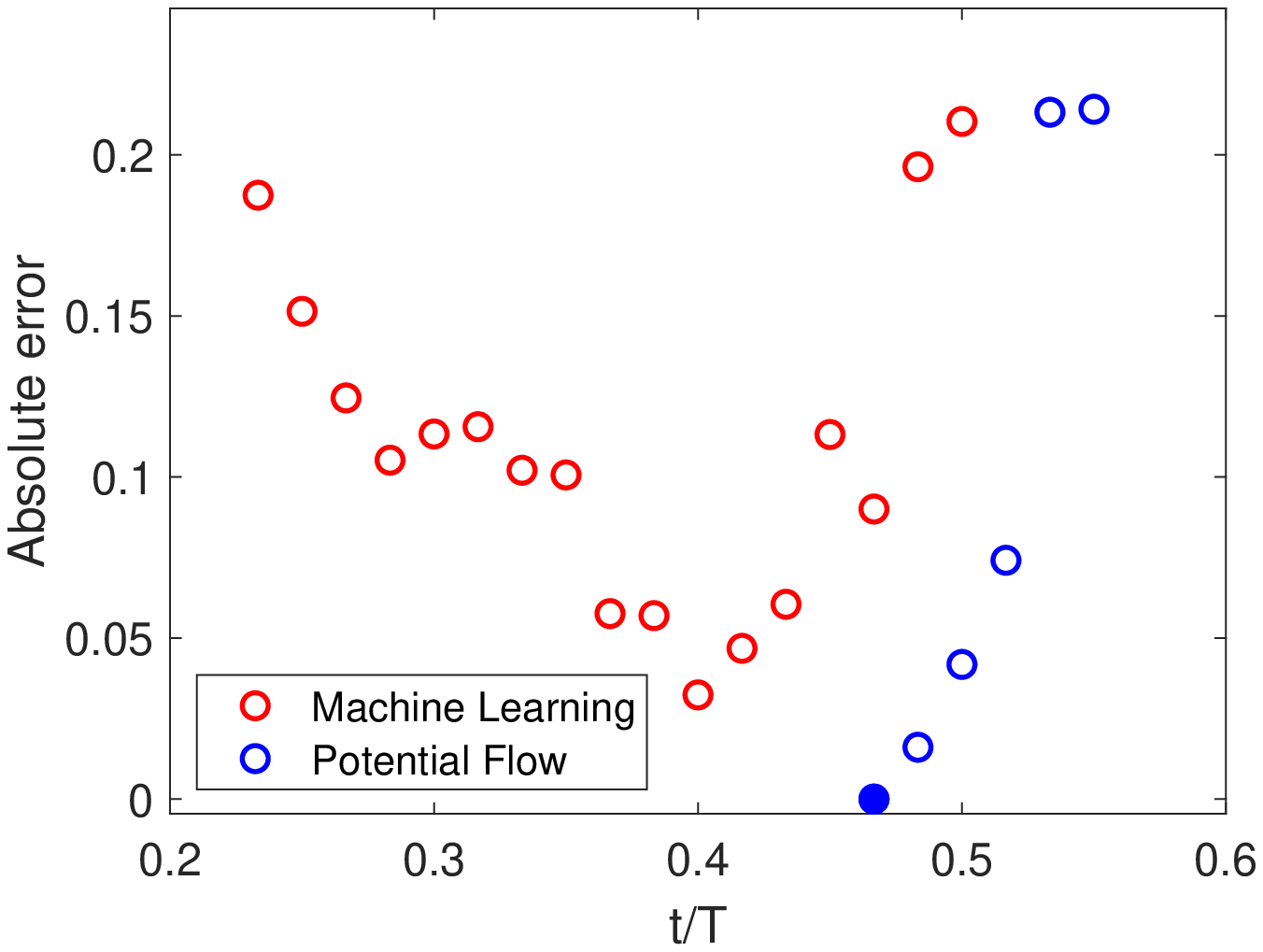}}
\subfloat[Case iv]{\label{main:d}\includegraphics[scale=.4]{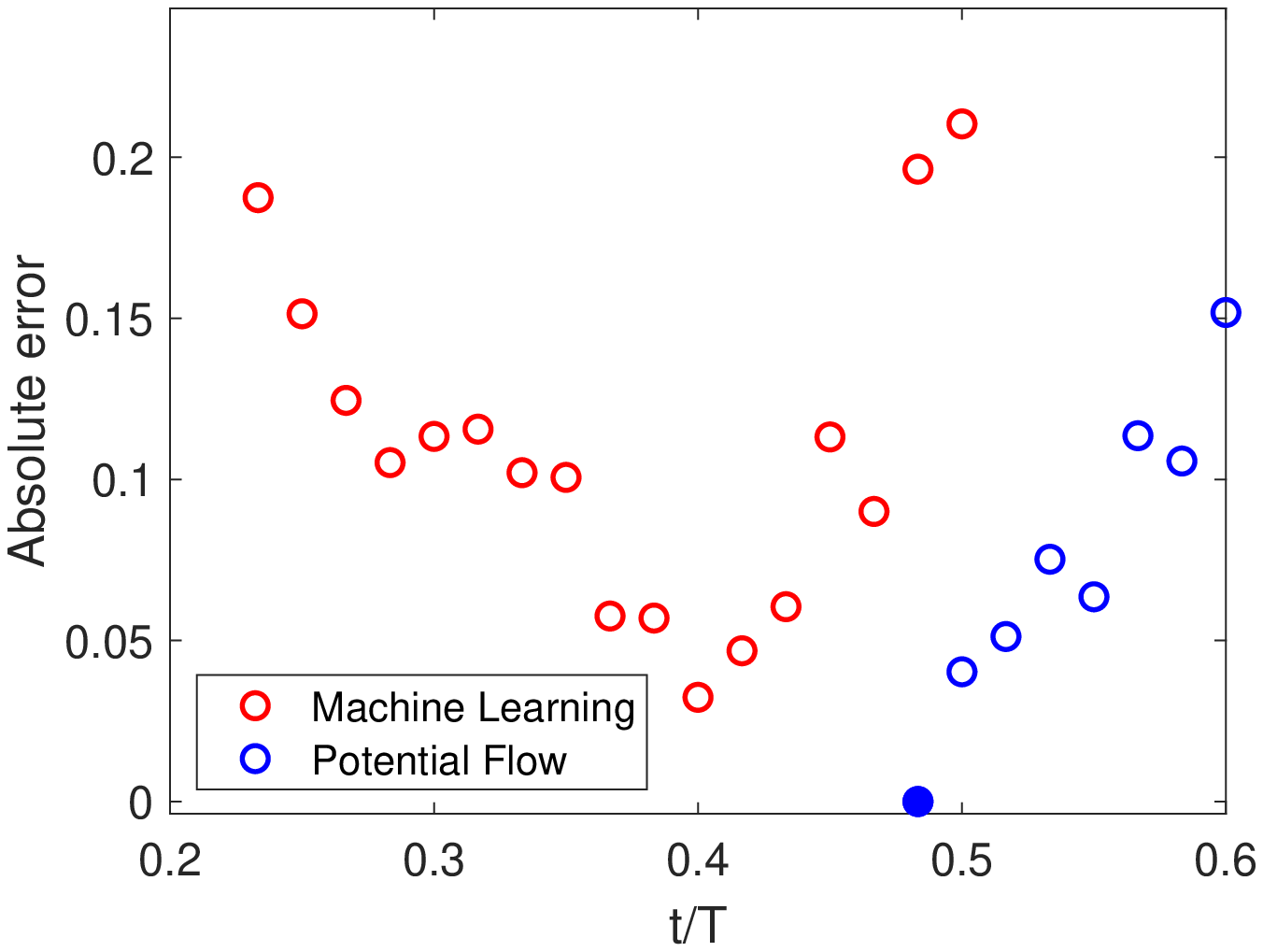}}
\caption{The position error, defined by the equation $\sqrt{(x_0 - x)^2 + (y_0 - y)^2}$, of the LEV trajectory predicted by machine learning and potential flow models compared to the measured trajectory. The solid blue marker indicates the time at which the potential flow model is initiated,  and the error is thus zero. $x_0, y_0$ are the predicted x,y positions of the LEV centroids while $x, y$ are the measured x, y positions. Refer to Table \ref{table:verificationParameters} for foil kinematics.}
\label{fig:errors}
\end{figure}